# Higher Variations of the Monty Hall Problem (3.0 and 4.0) and Empirical Definition of the Phenomenon of Mathematics, In Boole's Footsteps, as Something the Brain Does

**Leo Depuydt**[1], with an Appendix by **Richard D. Gill**[2]

[1]*Department of Egyptology and Ancient Western Asian Studies, Brown University, Providence, R. I., U.S.A.*
[1]*Department of the Classics, Harvard University, Cambridge, M. A., U.S.A. (Visiting Scholar)*
[2]*Mathematisch Instituut, Universiteit Leiden, The Netherlands*
E-mail: *leo_depuydt@brown.edu, gill@math.leidenuniv.nl*

**Abstract**

In *Advances in Pure Mathematics* (www.scirp.org/journal/apm), Vol. 1, No. 4 (July 2011), pp. 136-154, the mathematical structure of the much discussed problem of probability known as the Monty Hall problem was mapped in detail. It is styled here as Monty Hall 1.0. The proposed analysis was then generalized to related cases involving any number of doors ($d$), cars ($c$), and opened doors ($o$) (Monty Hall 2.0) and 1 specific case involving more than 1 picked door ($p$) (Monty Hall 3.0). In cognitive terms, this analysis was interpreted in function of the presumed digital nature of rational thought and language.

In the present paper, Monty Hall 1.0 and 2.0 are briefly reviewed (§§2-3). Additional generalizations of the problem are then presented in §§4-7. They concern expansions of the problem to the following items: (1) to any number of picked doors, with $p$ denoting the number of doors initially picked and $q$ the number of doors picked when switching doors after doors have been opened to reveal goats (Monty Hall 3.0; see §4); (3) to the precise conditions under which one's chances increase or decrease in instances of Monty Hall 3.0 (Monty Hall 3.2; see §6); and (4) to any number of switches of doors ($s$) (Monty Hall 4.0; see §7). The afore-mentioned article in *APM*, Vol. 1, No. 4 may serve as a useful introduction to the analysis of the higher variations of the Monty Hall problem offered in the present article.

An appendix by Richard D. Gill (see §8) provides additional context by building a bridge to modern probability theory in its conventional notation and by pointing to the benefits of certain interesting and relevant tools of computation now available on the Internet.

The cognitive component of the earlier investigation is extended in §9 by reflections on the foundations of mathematics. It will be proposed, in the footsteps of George Boole, that the phenomenon of mathematics needs to be defined in empirical terms as something that happens to the brain or something that the brain does. It is generally assumed that mathematics is a property of nature or reality or whatever one may call it. There is not the slightest intention in this paper to falsify this assumption because it cannot be falsified, just as it cannot be empirically or positively proven. But there is no way that this assumption can be a factual observation. It can be no more than an altogether reasonable, yet fully secondary, inference derived mainly from the fact that mathematics appears to work, even if some may deem the fact of this match to constitute proof. On the deepest empirical level, mathematics can only be directly observed and therefore directly analyzed as an activity of the brain. The study of mathematics therefore becomes an essential part of the study of cognition and human intelligence. The reflections on mathematics as a phenomenon offered in the present article will serve as a prelude to planned articles on how to redefine the foundations of probability as one type of mathematics in cognitive fashion and on how exactly Boole's theory of probability subsumes, supersedes, and completes classical probability theory.

§§2-7 combined, on the one hand, and §9, on the other hand, are both self-sufficient units and can be read independently from one another.

The ultimate design of the larger project of which this paper is part remains the increase of digitalization of the analysis of rational thought and language, that is, of (rational, not emotional) human intelligence. To reach out to other disciplines, an effort is made to describe the mathematics more explicitly than is usual.

**1. Introduction**

In *Advances in Pure Mathematics*, 2011, Vol. 1, No. 4, pp. 136-154, the mathematical structure of the well-known problem of probability known as the Monty Hall problem (see §2 below) was mapped in detail [1]. This mathematical structure includes two components that complement one another seamlessly. One component is digital or non-quantitative. The other is quantitative. The focus of that earlier paper was mainly on the neglected digital component. The digital component was analyzed in the spirit and the algebra of George Boole's *Investigation of the Laws of Thought* (1854), the Magna Charta of the digital age. Much of what has been said in the earlier paper is presupposed in what follows.

In said article, the analysis of the Monty Hall problem was extended in two directions. First, on the cognitive side, the digital analysis was interpreted as an organic reflection of the presumed digital nature of human cognition as expressed by rational thought and language and as evidenced empirically by facts of language. Probing the nature of rational thought and language was in a sense the ulterior motive of analyzing the Monty Hall problem. Second, on the mathematical side, the Monty Hall problem was generalized to related cases in accordance with the axioms of probability theory (Monty Hall 2.0). The aim was to demonstrate the reliability and productivity of the proposed digital approach. This first generalization is briefly reviewed in §3 below.

The analysis of the Monty Hall problem is extended again, both mathematically and cognitively, in the present paper. First, in mathematical terms, the validity of the proposed digital approach is bolstered by additional generalizations of the Monty Hall problem in '4, §5, §6, and §7 (Monty Hall 3.0 and 4.0). This process could presumably be carried on *ad infinitum*, at some point entering the domain of calculus.

Second, in cognitive terms, an attempt is made to render the presumed deep organic link between the digital component of probability theory and the digital nature of rational thought and language more probable by defining what mathematics is (see §8 below). In terms of the search for the deepest foundations of mathematics, it is proposed that mathematics is best defined first and foremost as something that the brain does as it engages reality outside itself through the senses.

§§2-7 combined, on the one hand, and §9, on the other hand, are self-sufficient and can be read independently from one another. In other words, it is not necessary to read §§2-7 in order to read §9.

An appendix by Richard D. Gill (see §8) provides additional context by building a bridge to modern probability theory in its conventional notation and by pointing to the benefits of certain interesting and relevant tools of computation now available on the Internet.

It is hoped that the reflections presented in §9 on the nature and definition of mathematics will serve as a prelude to forthcoming papers on the foundations of probability theory as one type of mathematics entitled "How Boole's Theory of Probability Subsumes, Supersedes, and Completes Classical Probability Theory: A Digital, Quantitative, and Cognitive Analysis," in which an attempt will be made to describe how exactly Boole's theory of probability, which has been almost entirely neglected for one and a half centuries, makes the classical theory of probability complete. It is imperative that a mathematical theory consider all possible cases. Classical probability theory does not.

H.H. Goldstine writes about Boole that "our debt to this simple, quiet man . . . is extraordinarily great and probably not adequately repaid" [2]. Goldstine is referring to the enormous significance of Boole's digital mathematics in modern computer science. It is suggested in §8 that the extent of the debt may far exceed computer science and reach deeply into the analysis of rational thought and language or human intelligence.

## 2. Monty Hall 1.0: The Original Monty Hall Problem, Featuring 1 Car (*c*), 3 Doors (*d*), 1 Opened Door (*o*), 1 Door Initially Picked (*p*), and 1 Door Picked by Switching (*q*)

Behind 3 closed doors, 2 goats and 1 car are hiding. One picks 1 door with the aim of getting the 1 car. The 1 door that one picks remains closed, however. Next, someone who knows what is hiding behind all the doors opens 1 of the 2 doors that were not picked, more specifically 1 door hiding a goat. 2 doors remain closed and available for picking, including the one initially picked. The Monty Hall problem involves the following question: Should one switch from the unopened door that one initially picked to the other door that remains unopened to improve one's chances of getting the car? The answer is: One should, because one doubles one's chances of getting the car—namely from 1 in 3 to 2 in 3—by switching doors once 1 door has been opened to reveal 1 goat.

## 3. Monty Hall 2.0: Generalization to Any Number of Doors (*d*), Cars (*c*), and Opened Doors (*o*)

The present generalization is treated in detail in the article mentioned in §1 above. What follows is a brief summary of this treatment.

The Monty Hall problem involves 1 car (*c*), 2 goats (*g*), 3 doors (*d*), 1 opened door (*o*), and 1 picked door (*p*). There are 5 variables. But in extending and generalizing the Monty Hall problem, only 4 variables need to be considered. That is because, of the 3 variables *c*, *g*, and *d*, each can be derived from the two others. From the fact that
$$c + g = d,$$
it follows that
$$c = d - g \quad \text{and} \quad g = d - c.$$
Only 2 of the variables *c*, *g*, and *d* therefore need to be considered. In what follows, *c* (cars)

and *d* (doors) are chosen.

As a general rule, in Monty Hall 2.0, one *always* improves one's chances of getting a car by switching doors when doors are opened to reveal goats. This will no longer be the case from Monty Hall 3.0 onward (see §4.12 and §6). The question remains: By how much? If the Monty Hall problem is generalized to any number of cars (*c*), doors (*d*), and opened doors (*o*), and only 1 door is picked, the chance of getting the car (*C*) by switching (*s*) doors ($C_s$) is

$$\frac{c(d-1)}{d(d-1-o)} \quad (1),$$

and the factor by which one improves one's chances of getting the car by switching is

$$\frac{d-1}{d-1-o} \quad (2).$$

The number 1 in these expressions represents the number of picked doors (*p*), which is fixed at 1.

For example, let there be 123,456,789 (or more than 123 million) doors (*d*), of which 12,345,678 (or more than 12.3 million) hide cars (*c*). Also assume that 1,234,567 (or more than 1.23 million) doors are opened (*o*) to reveal goats. The chances of getting a car (*C*) by switching (*s*) doors ($C_s$) is, according to expression (1),

$$\frac{12,345,678(123,456,789-1)}{123,456,789(123,456,789-1-1,234,567)} = \text{about } 0.101 \text{ or } 10.1\%.$$

The factor by which one *increases* one's chances of getting a car by switching doors is, according to expression (2),

$$\frac{123,456,789-1}{123,456,789-1-1,234,567} = \text{about } 1.010.$$

If this factor were 1, one would not increase one's chances because multiplying any number by 1 does not increase that number. But because the factor is about 1.010, one increases one's chances by about by about 0.01 or about 1%.

One's chances of getting a car when initially picking 1 door is the fraction of which the number of cars (*c*) is the numerator and the number of doors (*d*) the denominator, namely *c/d*, which in this case is

$$\frac{12,345,678}{123,456,789} = \text{about } 0.0999999927 \text{ or just about } 10\%.$$

Increasing one's chances from about 10% to about 10.1% indeed involves an increase of 1%, since 1% of 10 is about 0.1.

Since there are 123,456,789 doors (*d*) and 12,345,678 cars (*c*), there are 111,111,111 goats (*g*). According to the rules of the extended Monty Hall problem, up to $g-1$ doors can be opened to reveal goats, that is, 111,111,110 doors can be opened (*o*). If one opens the maximum number of doors that one is allowed to open, then according to expression (2) one increases one's chances of getting a car by switching by a factor of

$$\frac{123,456,789-1}{123,456,789-1-111,111,110} = 12,345,678.$$

Since a factor of 1 corresponds to a 0% increase, a factor of 2 to a 100% increase, a factor of 3 to 200% increase, and so on, a factor of 12,345,678 corresponds to an increase of 1,234,567,700%. In other words, one improves one's chances of getting a car by more than 1.23 billion percent by switching.

As regards the basic treatment of the Monty Hall problem in the afore-mentioned article, an additional note on notation is in order. Boole never ceased to impress upon his readers that probability is a field of mathematics that straddles the digital-mathematical and the quantitative-mathematical. The digital-mathematical and the quantitative-mathematical coexist in the single phenomenon of probability. To use a metaphor, it is a bit like Christianity's Trinity, three divine entities coexisting as one, although in this case not a trinity but a Duality is concerned. In probability as a field of mathematics, the digital-mathematical and the quantitative-mathematical are two facets of what is ultimately a single thing. Naturally, the human brain cannot quite think about the two facets at the very same time. But that is just a limitation of our mental capacities.

In Boole's notation, this coexistence of two facets in a single phenomenon is evoked felicitously by the single symbol $\times$ admitting of two interpretations. Consider the following two equivalent expressions found in the afore-mentioned article [3]:

$$C_i \times C_s : \frac{c}{d} \times \frac{g-o}{d-1-o}.$$

Both expressions describe the probability of initially picking a car and then picking a goat or non-car by switching.

The expression to the left of the colon is digital-mathematical. In this expression, the quantitative aspect is irrelevant. Accordingly, the symbols $C_i$ and $C_s$ are not quantitative. Likewise, if one divides the universe in strictly digital terms into four digital combination classes involving the two classes "black" ($b$) and "cat" ($c$), then the universe (1) equals $bc + \overline{b}c + b\overline{c} + \overline{b}\,\overline{c}$, that is, black cats, non-black cats, black things that are not cats, and things that are neither black nor cats. The sets $bc$ "black cats" and $\overline{b}c$ "non-black cats" will in all probability differ in quantity, assuming that it is possible to count all black and non-black cats. However, the difference in quantity is irrelevant in the digital-mathematical expression of the universe.

The expression to the right of the colon is quantitative-mathematical. Indeed, the symbols $c$, $g$, $d$, and $o$ are quantitative. They stand for numbers of cars, goats, doors, and opened doors. It follows that the symbol $\times$ admits of both a digital and a quantitative interpretation. The two interpretations may denoted by $\times_d$ and $\times_q$. Accordingly, the following equation applies:

$$C_i \times_d C_s \text{ (digital)} = \frac{c}{d} \times_q \frac{g-o}{d-1-o} \text{ (quantitative)}.$$

Multiplication is commutative. That means that $a \times b = b \times a$. However, it may be tempting to assume that $\times_d$ is not commutative. It is a fact that the event $C_s$, not getting a car by switching doors, *follows* the event $C_i$, initially getting a car, *in time*. And yet, in contemplating the combination of $C_i$ and $C_s$, nothing prevents one from contemplating $C_s$ first. The order in which one contemplates the two does not matter mathematically, even if it may come more naturally to think first of what comes first in time. Likewise, on the quantitative-mathematical level, the following equation applies:

$$\frac{c}{d} \times_q \frac{g-o}{d-1-o} = \frac{g-o}{d-1-o} \times_q \frac{c}{d}.$$

**4. Monty Hall 3.0: Additional Generalization to Any Number of Doors Picked Initially ($p$) or of Doors Picked by Switching ($q$)**

**4.1. The Special Case of Getting *at Least 1 Car* when Switching Doors**

In Monty Hall 1.0 and 2.0, just 1 door is picked both before and after switching. The most natural expansion of 1.0 and 2.0 would seem to be the generalization in which any number of doors are picked both before and after switching. The number of doors picked will be denoted by $p$; the number of doors picked by switching, by $q$. The present generalization is styled here as Monty Hall 3.0.

One can imagine many desired outcomes of picking 1 or more doors. For example, the desired outcome might be to get cars with every door pick both before and after switching. Or the desired outcome might be to obtain 1 car in the 1$^{st}$ and the 3$^{rd}$ of 3 initial picks as compared to picking 1 car in the 1$^{st}$ of 2 picks by switching. And so on. Treating all desired outcomes comprehensively exceeds the scope of the present paper. In such a comprehensive treatment, it is necessary to take one's departure from the equation representing the total probability of all possible outcomes, whose individual probabilities add up to 1 or 100%. It is hoped that it will be possible to present a survey of the respective probabilities of all possible outcomes in a future paper.

Presently, just 1 desired outcome will be selected. The aim is to select an outcome that concords with the spirit of the original Monty Hall problem. In the original problem, the person picking a door wants a car. Accordingly, when more than 1 door is picked, the desired outcome that most closely reflects the spirit of the original Monty Hall problem is getting at least 1 car. It would be awkward to deny the person any car at all if more than 1 car are picked.

The probability $P$ that one will get at least 1 car by switching doors is a fraction whose numerator is $N$ and whose denominator is $D$. $N$ and $D$ are defined below. Most the rest of §4 is devoted to a description of how the equation below is obtained. A more explicit version of the numerator appears in §4.17 below.

The precise relation between the following expression and the common probabilistic conceptualization known as hypergeometric distribution will be described in a future paper.

$$P = \frac{N}{D}$$

[*Editors and typesetters, please watch the tabs/the indentation and delete this line after editing!*]

$$N = \frac{c!}{(c-p)!}$$
$$\times [\frac{(c-p)!}{(c-p-q)!} + q\frac{(c-p)!}{[c-p-(q-1)]!}$$
$$\times \frac{[g-o-(p-p)]!}{\{g-o-(p-p)-[q-(q-1)]\}!}$$
$$+ \frac{q(q-1)}{1 \times 2} \times \frac{(c-p)!}{[c-p-(q-2)]!}$$
$$\times \frac{[g-o-(p-p)]!}{\{g-o-(p-p)-[q-(q-2)]\}!}$$
$$+ \frac{q(q-1)(q-2)}{1 \times 2 \times 3} \times \frac{(c-p)!}{[c-p-(q-3)]!}$$
$$\times \frac{[g-o-(p-p)]!}{\{g-o-(p-p)-[q-(q-3)]\}!}$$

$$+ \ldots + \frac{q(q-1)(q-2)\times\ldots\times[q-(q-3)][q-(q-2)]}{1\times 2\times 3\times\ldots\times(q-2)(q-1)}$$

$$\times \frac{(c-p)!}{[c-p-[q-(q-1)]]!}$$

$$\times \frac{[g-o-(p-p)]!}{\langle g-o-(p-p)-\{q-[q-(q-1)]\}\rangle!} ]$$

$$+ p\frac{c!}{[c-(p-1)]!} \times \frac{g!}{\{g-[p-(p-1)]\}!}$$

$$\times [\frac{[c-(p-1)]!}{[c-(p-1)-q)]!}$$

$$\times \frac{\{g-o-[p-(p-1)]\}!}{\{g-o-[p-(p-1)]-(q-q)\}!}$$

$$+ q\frac{[c-(p-1)]!}{[c-(p-1)-(q-1)]!}$$

$$\times \frac{\{g-o-[p-(p-1)]\}!}{\{g-o-[p-(p-1)]-[q-(q-1)]\}!}$$

$$+ \frac{q(q-1)}{1\times 2} \times \frac{[c-(p-1)]!}{[c-(p-1)-(q-2)]!}$$

$$\times \frac{\{g-o-[p-(p-1)]\}!}{\{g-o-[p-(p-1)]-[q-(q-2)]\}!}$$

$$+ \frac{q(q-1)(q-2)}{1\times 2\times 3} \times \frac{[c-(p-1)]!}{[c-(p-1)-(q-3)]!}$$

$$\times \frac{\{g-o-[p-(p-1)]\}!}{\{g-o-[p-(p-1)]-[q-(q-3)]\}!}$$

$$+ \ldots + \frac{q(q-1)(q-2)\times\ldots\times[q-(q-3)][q-(q-2)]}{1\times 2\times 3\times\ldots\times(q-2)(q-1)}$$

$$\times \frac{[c-(p-1)]!}{\{c-(p-1)-[q-(q-1)]\}!}$$

$$\times \frac{\{g-o-[p-(p-1)]\}!}{\langle\{g-o-[p-(p-1)]-\{q-[q-(q-1)]\}\rangle!} ]$$

$$+ \frac{p(p-1)}{1\times 2} \times \frac{c!}{[c-(p-2)]!} \times \frac{g!}{\{g-[p-(p-2)]\}!}$$

$$\times [\frac{[c-(p-2)]!}{[c-(p-2)-q]!}$$

$$\times \frac{\{g-o-[p-(p-2)]\}!}{\{g-o-[p-(p-2)]-(q-q)\}!}$$

$$+ q\frac{[c-(p-2)]!}{[c-(p-2)-(q-1)]!}$$

$$\times \frac{\{g-o-[p-(p-2)]\}!}{\{g-o-[p-(p-2)]-[q-(q-1)]\}!}$$

$$+ \frac{q(q-1)}{1\times 2} \times \frac{[c-(p-2)]!}{[c-(p-2)-(q-2)]!}$$

$$\times \frac{\{g-o-[p-(p-2)]\}!}{\{g-o-[p-(p-2)]-[q-(q-2)]\}!}$$

$$+ \frac{q(q-1)(q-2)}{1\times 2\times 3} \times \frac{[c-(p-2)]!}{[c-(p-2)-(q-3)]!}$$

$$\times \frac{\{g-o-[p-(p-2)]\}!}{\{g-o-[p-(p-2)]-[q-(q-3)]\}!}$$

$$+\ldots + \frac{q(q-1)(q-2)\times\ldots\times[q-(q-3)][q-(q-2)]}{1\times 2\times 3\times\ldots\times(q-2)(q-1)}$$

$$\times \frac{[c-(p-2)]!}{\{c-(p-2)-[q-(q-1)]\}!}$$

$$\times \frac{\{g-o-[p-(p-2)]\}!}{\langle g-o-[p-(p-2)]-\{q-[q-(q-1)]\}\rangle!}\,]$$

$$+ \frac{p(p-1)(p-2)}{1\times 2\times 3} \times \frac{c!}{[c-(p-3)]!}$$

$$\times \frac{g!}{\{g-[p-(p-3)]\}!}$$

$$\times [\,\frac{[c-(p-3)]!}{[c-(p-3)-q]!}$$

$$\times \frac{\{g-o-[p-(p-3)]\}!}{\{g-o-[p-(p-3)]-(q-q)\}!}$$

$$+ q\frac{[c-(p-3)]!}{[c-(p-3)-(q-1)]!}$$

$$\times \frac{\{g-o-[p-(p-3)]\}!}{\{g-o-[p-(p-3)]-[q-(q-1)]\}!}$$

$$+ \frac{q(q-1)}{1\times 2} \times \frac{[c-(p-3)]!}{[c-(p-3)-(q-2)]!}$$

$$\times \frac{\{g-o-[p-(p-3)]\}!}{\{g-o-[p-(p-3)]-[q-(q-2)]\}!}$$

$$+ \frac{q(q-1)(q-2)}{1\times 2\times 3} \times \frac{[c-(p-3)]!}{[c-(p-3)-(q-3)]!}$$

$$\times \frac{\{g-o-[p-(p-3)]\}!}{\{g-o-[p-(p-3)]-[q-(q-3)]\}!}$$

$$+\ldots + \frac{q(q-1)(q-2)\times\ldots\times[q-(q-3)][q-(q-2)]}{1\times 2\times 3\times\ldots\times(q-2)(q-1)}$$

$$\times \frac{[c-(p-3)]!}{\{c-(p-3)-[q-(q-1)]\}!}$$

$$\times \frac{\{g-o-[p-(p-2)]\}!}{\langle g-o-[p-(p-2)]-\{q-[q-(q-1)]\}\rangle!}\,]$$

$$+\ldots + \frac{p(p-1)(p-2)\times\ldots\times[p-p-2][p-(p-1)]}{1\times 2\times 3\times\ldots\times(p-1)p}$$

$$\times \frac{c!}{[c-(p-p)]!} \times \frac{g!}{\{g-[p-(p-p)]\}!}$$

$$\times [\frac{[c-(p-p)]!}{[c-(p-p)-q]!}$$

$$\times \frac{\{g-o-[p-(p-p)]\}!}{\{g-o-[p-(p-p)]-(q-q)\}!}$$

$$+ q \frac{[c-(p-p)]!}{[c-(p-p)-(q-1)]!}$$

$$\times \frac{\{g-o-[p-(p-p)]\}!}{\{g-o-[p-(p-p)]-[q-(q-1)]\}!}$$

$$+ \frac{q(q-1)}{1 \times 2} \times \frac{[c-(p-p)]!}{[c-(p-p)-(q-2)]!}$$

$$\times \frac{\{g-o-[p-(p-p)]\}!}{\{g-o-[p-(p-p)]-[q-(q-2)]\}!}$$

$$+ \frac{q(q-1)(q-2)}{1 \times 2 \times 3} \times \frac{[c-(p-p)]!}{[c-(p-p)-(q-3)]!}$$

$$\times \frac{\{g-o-[p-(p-p)]\}!}{\{g-o-[p-(p-p)]-[q-(q-3)]\}!}$$

$$+ \ldots + \frac{q(q-1)(q-2) \times \ldots \times [q-(q-3)][q-(q-2)]}{1 \times 2 \times 3 \times \ldots \times (q-2)(q-1)}$$

$$\times \frac{[c-(p-p)]!}{\{c-(p-p)-[q-(q-1)]\}!}$$

$$\times \frac{\{g-o-[p-(p-p)]\}!}{\langle g-o-[p-(p-p)]-\{q-[q-(q-1)]\}\rangle!} ]$$

$$D = \frac{d!}{(d-p)!} \times \frac{(d-p-o)!}{(d-p-o-q)!}.$$

In the following maximally compacted version of the numerator, the coefficient terms have been reduced from 5 to 4 and much of the transparency has been lost.

**[Editors and typesetters, please watch the tabs/the indentation and delete this line after editing!]**

$$N = \frac{c!}{(c-p)!}$$

$$\times [\frac{(c-p)!}{(c-p-q)!} + q \frac{(c-p)!}{(c-p-q+1)!} \times \frac{(g-o)!}{(g-o-1)!}$$

$$+ \frac{(c-p)!}{(c-p-q+2)!} \times \frac{(g-o)!}{(g-o-2)!}$$

$$+ \ldots + \frac{q(q-1)(q-2) \times \ldots \times 2}{1 \times 2 \times 3 \times \ldots \times (q-1)}$$

$$\times \frac{(c-p)!}{(c-p-1)!} \times \frac{(g-o)!}{(g-o-q-1)!} ]$$

$$+ p \frac{c!}{(c-p+1)!} \times \frac{g!}{(g-1)!}$$

$$\times [\frac{(c-p+1)!}{(c-p-q+1)!} + q \frac{(c-p+1)!}{(c-p-q+2)!}$$

$$\times \frac{(g-o-1)!}{(g-o-2)!}$$

$$+ \frac{q(q-1)}{1 \times 2} \times \frac{(c-p+1)!}{(c-p-q+3)!} \times \frac{(g-o-1))!}{(g-o-3)!}$$

$$+ \ldots + \frac{q(q-1)(q-2) \times \ldots \times 2}{1 \times 2 \times 3 \times \ldots \times (q-1)}$$

$$\times \frac{(c-p+1)!}{(c-p)!} \times \frac{(g-o-1)!}{(g-o-q)!}]$$

$$+ \frac{p(p-1)}{1 \times 2} \times \frac{c!}{(c-p+2)!} \times \frac{g!}{(g-2)!}$$

$$\times [\frac{(c-p+2)!}{(c-p-q+2)!} + q\frac{(c-p+2)!}{(c-p-q+3)!}$$

$$\times \frac{(g-o-2)!}{(g-o-3)!}$$

$$+ \frac{q(q-1)}{1 \times 2} \times \frac{(c-p+2)!}{(c-p-q+4)!} \times \frac{(g-o-2)!}{(g-o-4)!}$$

$$+ \ldots + \frac{q(q-1)(q-2) \times \ldots \times 2}{1 \times 2 \times 3 \times \ldots \times (q-1)}$$

$$\times \frac{(c-p+2)!}{(c-p+1)!} \times \frac{(g-o-2)!}{(g-o-q-1)!}]$$

$$+ \ldots + \frac{p(p-1)(p-2) \times \ldots \times 1}{1 \times 2 \times 3 \times \ldots \times p} \times \frac{g!}{(g-p)!}$$

$$\times [\frac{c!}{(c-q)!} + q\frac{c!}{(c-q+1)!} \times \frac{(g-o-p)!}{(g-o-p-1)!}$$

$$+ \frac{q(q-1)}{1 \times 2} \times \frac{c!}{(c-q+2)!} \times \frac{(g-o-p)!}{(g-o-p-2)!}$$

$$+ \frac{q(q-1)(q-2)}{1 \times 2 \times 3} \times \frac{c!}{(c-q+3)!}$$

$$\times \frac{(g-o-p)!}{(g-o-p-3)!}$$

$$+ \ldots + \frac{q(q-1)(q-2) \times \ldots \times 2}{1 \times 2 \times 3 \times \ldots \times (q-1)}$$

$$\times \frac{c!}{(c-1)!} \times \frac{(g-o-p)!}{(g-o-p-q+1)!}]$$

### 4.2. Point of Departure: An Example

It will be convenient to begin the description of how the Equation in §4.1 is obtained with a specific example. Once it is seen how the probability of getting at least 1 car by switching doors after doors hiding goats have been opened is obtained in 1 case, the result can be generalized to all cases. In the example that will be used here, the numbers of the variables are as follows:

    cars ($c$)                                 =  5;  
    goats ($g$)                                =  7;

doors ($d = c + g$) = 12;
doors picked initially ($p$) = 3;
doors subsequently opened ($o$) = 2;
doors picked by switching ($q$) = 2.

### 4.3. All Possible Scenarios as Sequences of 5 Picks of Doors Hiding Cars or Goats

It lies in the nature of probability that a number of different scenarios can be expected, each with its own degree of probability. In the example at hand, all possible scenarios consist of 5 successive picks of doors that hide either cars or goats. There are 3 initial picks of doors ($p = 3$) and 2 additional picks of doors by switching doors ($q = 2$) after 2 doors revealing goats have been opened ($o = 2$).

### 4.4. The 32 Possible Sequences of Picks

To be determined first are all the possible sequences of 5 picks in which either cars ($c$) or goats ($g$) are picked. There are 32 possible sequences, as follows: (1) *ccccc* (picking a car at every pick); (2) *ccccg* (picking 4 cars and then 1 goat); (3) *cccgc* (picking 3 cars, then 1 goat, and finally 1 car); (4) *cccgg*; (5) *ccgcc*; (6) *ccgcg*; (7) *ccggc*; (8) *ccggg*; (9) *cgccc*; (10) *cgccg*; (11) *cgcgc*; (12) *cgcgg*; (13) *gcccc*; (14) *gcccg*; (15) *gccgc*; (16) *gccgg*; (17) *cggcc*; (18) *cggcg*; (19) *cgggc*; (20) *cgggg*; (21) *gcgcc*; (22) *gcgcg*; (23) *gcggc*; (24) *gcggg*; (25) *ggccc*; (26) *ggccg*; (27) *ggcgc*; (28) *ggcgg*; (29) *gggcc*; (30) *gggcg*; (31) *ggggc*; and (32) *ggggg*.

More generally speaking, the number 32 is obtained as follows according to the theory of permutations. At 1st pick, there are only 2 possible scenarios: one picks either a car or a goat. In the 1st and 2nd picks combined, there are 4 possible scenarios: after picking a car in the 1st pick, one can pick either a car or a goat in the 2nd pick; likewise, after picking a goat in the 1st pick, one can pick either a car or goat in the 2nd pick. In other words, the number of possible scenarios has doubled from the 1st pick to the 2nd pick from 2 or $2^1$ to 4 or $2^2$. It is easily seen that the number of possible scenarios will likewise double at every successive pick. Accordingly, the number of possible scenarios after 5 picks will be $2^5$ or 32.

### 4.5. The Changing Probabilities of Each Successive Pick

The probability of each single pick is a fraction whose numerator is either the number of available cars or the number of available goats and whose denominator is the number of available doors. At 1st pick, all the cars, goats, and doors are still available for picking. Accordingly, in the example at hand, the chance of picking a car is $c/d$, that is, 5/12, and the chance of picking a goat is $g/d$, that is, 7/12.

After each pick, the denominator or the number of doors decreases by 1, from $d$ to $d-1$, and so on. The number of available doors decreases additionally when doors are opened to reveal goats. In the example at hand, the number of available doors first decreases from 12 to 9 as 3 doors are picked. The number then further decreases to 7 when 2 doors are opened to reveal goats. Finally, the number decreases to 5 as 2 more doors are picked by switching doors.

The number of available cars does not decrease when a goat is picked. Nor does the number of available goats when a car is picked. By contrast, the number of doors decreases at *every* pick. It follows that the probability of picking a car or a goat changes at *every* successive pick because at least the number of available doors, which constitutes the denominator of the probability of each pick, changes.

### 4.6. Conditional Probability as a Property of All Picks Preceded by Other Picks

Each pick decreases the number of the available doors as well as either the number of available cars or the number of available goats. In that regard, each pick of either a car or a goat changes the probability of later picks of either a car or a goat. In other words, the probability of a later pick of either a goat or a car is *dependent* on what happens in an earlier pick or earlier picks. An event whose probability is affected by what happens in earlier events is called a dependent event. Events on which other events are dependent may be called lead events. In the Monty Hall problem and its extensions, only the very 1st car picks and the 1st goat picks of sequences of picks are *not* dependent. An event is usually called dependent in the context of the combined probability of 2 or more events in which some events are dependent and others are not. Thus, the combined probability of picking two cars in a row is $c/d \times (c-1)/(d-1)$. The 1st pick is the lead pick. The 2nd pick is the dependent pick.

Earlier picks serve as conditions of the probability of later picks. Accordingly, the general phenomenon in which the probability of a later event is changed by an earlier event from what its probability would have been without that earlier event taking place is called conditional probability.

For example, the probability of picking a car when all cars and all doors are still available is $c/d$. But once 1 car is picked, the number of cars and doors both decrease by 1, the assumption being that one cannot pick the same door twice. The probability of picking a car therefore changes to $(c-1)/(d-1)$. When a goat is picked instead of a car, the probability of picking a car changes instead to $c/(d-1)$. At the same time, the probability of picking a goat changes to $(g-1)/(d-1)$.

An event is usually called dependent in the context of the combined probability of two or more events in which some events are dependent and others are not. Thus, the combined probability of picking two cars in a row is $c/d \times (c-1)/(d-1)$. The 1st pick is the lead pick. The 2nd pick is the dependent pick.

The degree to which a prior event changes the probability of an event from what it would have been without that prior event can be quantified. In the example at hand, the change in probability from $c/d$ to $(c-1)/(d-1)$ that results from the pick of a car corresponds to a *diminution* in probability of 7/132, or about 5.3%, from 5/12 to 4/11. By contrast, the change from $c/d$ to $c/(d-1)$ that results from the pick of a goat corresponds to an *increase* in probability of 5/132, or about 3.8%, from 5/12 to 5/11. In sum, conditional probability is best *measured* or quantified as the degree of change between a 1st event and a 2nd event whose probability depends on the 1st event.

### 4.7. The General Denominator of the Equation in §4.1

It has been noted in §4.4 that there are 32 possible sequences of 5 picks in the example at hand. Each sequence of 5 picks comes with its own probability. The specific denominator of all 32 probabilities is the same, namely $d(d-1)(d-2) \times (d-o-3)(d-o-4)$, or in the example at hand, $12(12-1)(12-2) \times (12-2-3)(12-2-4)$, or $12 \times 11 \times 10 \times 7 \times 6$. In other words, $12 \times 11 \times 10 \times 7 \times 6$ is the common denominator of all 32 probabilities. The sign × separates the picks of doors before doors are opened from the picks of doors after doors are opened.

What is the general form of the denominator? It appears that the expression $12 \times 11 \times 10$ is the 1st portion of 12!, or of 12 factorial, or also of $12 \times 11 \times 10 \times 9 \times 8 \times 7 \times 6 \times 5 \times 4 \times 3 \times 2 \times 1$, and that $7 \times 6$ is the 1st portion of 7!, or of 7 factorial, or also of $7 \times 6 \times 5 \times 4 \times 3 \times 2 \times 1$. In fact, the two components of the denominator will always be portions of factorials. The need therefore arises to represent the two components of the denominator in general forms as portions of factorials.

In that regard, $12 \times 11 \times 10$ is nothing but 12! divided by 9!, or $12 \times 11 \times 10 \times 9 \times 8 \times 7 \times 6 \times 5 \times 4 \times 3 \times 2 \times 1$ divided by $9 \times 8 \times 7 \times 6 \times 5 \times 4 \times 3 \times 2 \times 1$. The elimination of the common factor $9 \times 8 \times 7 \times 6 \times 5 \times 4 \times 3 \times 2 \times 1$ yields the desired $12 \times 11 \times 10$. Likewise, $7 \times 6$ is the same as 7! divided by 5!, or $7 \times 6 \times 5 \times 4 \times 3 \times 2 \times 1$ divided by $5 \times 4 \times 3 \times 2 \times 1$. The elimination of the common factor $5 \times 4 \times 3 \times 2 \times 1$ yields the desired $7 \times 6$.

In converting 12!/9! into a general form, it appears that 12 is the number of doors ($d$) and 9 is the number of the doors ($d$) minus the number of doors picked before doors are opened ($p$), that is, $d - p$. Consequently, the general equivalent of specific 12!/9! is $d!/(d-p)!$. In converting 7!/5! into a general form, it appears that 7 is the number of doors picked before doors are opened ($p$) minus the number of opened doors ($o$), that is, $d - p - o$, and that 5 is the number of doors picked before doors are opened ($p$) minus the number of opened doors ($o$) minus the number of doors picked after doors are opened, that is, $d - p - o - q$. Consequently, the general equivalent of specific 7!/5! is $(d-p-o)!/(d-p-o-q)!$.

It may be concluded that the general form of the denominator of the fraction that expresses the probability that 1 will get at least 1 car by switching doors for any number of $d$, $c$, $g$, $p$, $o$, or $q$ is as follows:

$$\frac{d!}{(d-p)!} \times \frac{(d-p-o)!}{(d-p-o-q)!}.$$

### 4.8. The Specific Numerators of the Probabilities of the 32 Sequences of Picks in the Example at Hand

It has been noted in §4.4 that there are 32 possible sequences of 5 picks in the example at hand. Each sequence of 5 picks comes with its own probability. Each of these 5 probabilities is expressed by its own fraction and each fraction has its own numerator. The numerators of the 5 individual probabilities of all the 32 scenarios are as follows, with × again separating the picks of doors before doors are opened from the picks of doors after doors are opened.

1. $ccc \times cc$: $c(c-1)(c-2) \times (c-3)(c-4)$
2. $ccc \times cg$: $c(c-1)(c-2) \times (c-3)(g-o)$
3. $ccc \times gc$: $c(c-1)(c-2) \times (g-o)(c-3)$
4. $ccc \times gg$: $c(c-1)(c-2) \times (g-o)(g-o-1)$
5. $ccg \times cc$: $c(c-1)g \times (c-2)(c-3)$
6. $ccg \times cg$: $c(c-1)g \times (c-2)(g-o-1)$
7. $ccg \times gc$: $c(c-1)g \times (g-o-1)(c-2)$
8. $ccg \times gg$: $c(c-1)g \times (g-o-1)(g-o-2)$

9.  $cgc \times cc$:  $cg(c-1) \times (c-2)(c-3)$
10. $cgc \times cg$:  $cg(c-1) \times (c-2)(g-o-1)$
11. $cgc \times gc$:  $cg(c-1) \times (g-o-1)(c-2)$
12. $cgc \times gg$:  $cg(c-1) \times (g-o-1)(g-o-2)$
13. $gcc \times cc$:  $gc(c-1) \times (c-2)(c-3)$
14. $gcc \times cg$:  $gc(c-1) \times (c-2)(g-o-1)$
15. $gcc \times gc$:  $gc(c-1) \times (g-o-1)(c-2)$
16. $gcc \times gg$:  $gc(c-1) \times (g-o-1)(g-o-2)$
17. $cgg \times cc$:  $cg(g-1) \times (c-1)(c-2)$
18. $cgg \times cg$:  $cg(g-1) \times (c-1)(g-o-2)$
19. $cgg \times gc$:  $cg(g-1) \times (g-o-2)(c-1)$
20. $cgg \times gg$:  $cg(g-1) \times (g-o-2)(g-o-3)$
21. $gcg \times cc$:  $gc(g-1) \times (c-1)(c-2)$
22. $gcg \times cg$:  $gc(g-1) \times (c-1)(g-o-2)$
23. $gcg \times gc$:  $gc(g-1) \times (g-o-2)(c-1)$
24. $gcg \times gg$:  $gc(g-1) \times (g-o-2)(g-o-3)$
25. $ggc \times cc$:  $g(g-1)c \times (c-1)(c-2)$
26. $ggc \times cg$:  $g(g-1)c \times (c-1)(g-o-2)$
27. $ggc \times gc$:  $g(g-1)c \times (g-o-2)(c-1)$
28. $ggc \times gg$:  $g(g-1)c \times (g-o-2)(g-o-3)$
29. $ggg \times cc$:  $g(g-1)(g-2) \times c(c-1)$
30. $ggg \times cg$:  $g(g-1)(g-2) \times c(g-o-3)$
31. $ggg \times gc$:  $g(g-1)(g-2) \times (g-o-3)c$
32. $ggg \times gg$:  $g(g-1)(g-2) \times (g-o-3)(g-o-4)$

I refrain from detailing how each single factor in the products is obtained. The principles of conditional probability have been explicated above. Suffice it to note that the number of the cars available for picking decreases by 1 every time a car gets picked. And so does the number of the goats. In addition, the number of the goats decreases by the number of opened doors.

It is not the case that the 32 sequences of picks are all equally probable. For example, picking 5 cars in a row (no. 1) is naturally less probable than picking 5 goats in a row (no. 32) because there are fewer cars to pick.

### 4.9. The Specific Numerators of the Probabilities of the 24 Sequences of Picks that Yield at least 1 Car in the Example at Hand

In order to obtain the numerator of the probability that 1 will get at least 1 car by switching doors, only those sequences of picks in which either or both of the 2 picks made after doors have been opened yield at least 1 car can be considered. Or, the 8 sequences that yield no car need to be eliminated. They are sequences 4, 8, 12, 16, 20, 24, 28, and 32. In the list below, the 8 sequences in question have been removed. What is more, the sequences have been reordered and so have the factors within the sequences to assimilate like to like. Reordering the factors is obviously possible because multiplication is commutative. But no factors have been moved across the symbol × because the factors at both sides of × belong to different picks as events. Also, the order of $c$ and $g$ has not been changed in the expressions of the type $ccc \times cc$. The result of this reordering is the following 8 groups of sequences, numbered i-viii.

Group i
1. $ccc \times cc$: $c(c-1)(c-2) \times (c-3)(c-4)$

Group ii
2. $ccc \times cg$: $c(c-1)(c-2) \times (c-3)(g-o)$
3. $ccc \times gc$ : $c(c-1)(c-2) \times (c-3)(g-o)$

Group iii
5. $ccg \times cc$: $c(c-1)g \times (c-2)(c-3)$
9. $cgc \times cc$: $c(c-1)g \times (c-2)(c-3)$
13. $gcc \times cc$: $c(c-1)g \times (c-2)(c-3)$

Group iv
6. $ccg \times cg$: $c(c-1)g \times (c-2)(g-o-1)$
7. $ccg \times gc$: $c(c-1)g \times (c-2)(g-o-1)$
10. $cgc \times cg$: $c(c-1)g \times (c-2)(g-o-1)$
11. $cgc \times gc$: $c(c-1)g \times (c-2)(g-o-1)$
14. $gcc \times cg$: $c(c-1)g \times (c-2)(g-o-1)$
15. $gcc \times gc$: $c(c-1)g \times (c-2)(g-o-1)$

Group v
17. $cgg \times cc$: $cg(g-1) \times (c-1)(c-2)$
21. $gcg \times cc$: $cg(g-1) \times (c-1)(c-2)$
25. $ggc \times cc$: $cg(g-1) \times (c-1)(c-2)$

Group vi
18. $cgg \times cg$: $cg(g-1) \times (c-1)(g-o-2)$
19. $cgg \times gc$: $cg(g-1) \times (c-1)(g-o-2)$
22. $gcg \times cg$: $cg(g-1) \times (c-1)(g-o-2)$
23. $gcg \times gc$: $cg(g-1) \times (c-1)(g-o-2)$
26. $ggc \times cg$: $cg(g-1) \times (c-1)(g-o-2)$
27. $ggc \times gc$: $cg(g-1) \times (c-1)(g-o-2)$

Group vii
29. $ggg \times cc$: $g(g-1)(g-2) \times c(c-1)$

Group viii
30. $ggg \times cg$: $g(g-1)(g-2) \times c(g-o-3)$
31. $ggg \times gc$: $g(g-1)(g-2) \times c(g-o-3)$

The design of what follows is to construct the general expression for the probability that 1 will get the car by switching for any number of $d$, $c$, $g$, $p$, $o$, or $q$ from the 24 products of 5 factors listed above. In doing so, I am deliberately more explicit than might otherwise be the case in a mathematics journal in order to be more accessible and inviting. The ulterior design of the present effort lies after all beyond mathematics. It is the description of the structure of human intelligence.

In turning the example at hand into a general expression, two operations need to be performed: (1) the addition of two sequences of coefficients, one relating to $p$ and the other to $q$, and (2) the addition of factorials. Once these two operations have been performed, it can be determined whether any simplifications are possible. The addition of factorials has already been discussed above. For example, a product such as $c(c-1)(c-2)$, that is, $5 \times 4 \times 3$ in the example at hand, can first be converted into $c!/(c-3)!$, that is, $5!/(5-3)!$ or

$5 \times 4 \times 3 \times 2 \times 1/2 \times 1$ in the example at hand. It can then be generalized to $c!/(c-p)!$. It will therefore be useful to turn first to the coefficients, which involve "the most famous of all number patterns" [4].

### 4.10. The Coefficients of the Probabilities of the 24 Sequences of Picks that Yield at least 1 Car in the Example at Hand

In the list of products in §4.9, in which like has been assimilated to like, there are 8 groups of sequences of picks consisting of 1, 2, 3, 6, 3, 6, 1, and 2 sequences respectively. How can these numbers be accounted for?

On closer inspection, it appears that they have everything to do with how many permutations of *c* and *g* there are in the 2 components before and after $\times$.

For example, in the 2nd group consisting of nos. 2 and 3, the initial product is either $ccc \times cg$ or $ccc \times gc$. Before $\times$, there is 1 permutation, namely *ccc*. After $\times$, there are 2 permutations, namely *cg* and *gc*. Accordingly, there are $1 \times 2$ or 2 members in the group.

In the 3rd group, there are 3 permutations before $\times$, namely *ccg*, *cgc*, and *gcc*, and 2 permutations after $\times$, namely *cg* and *gc*. Accordingly, there are $3 \times 2$ or 6 members in the group.

Furthermore, the reason that there are 3 permutations before the symbol $\times$ in the 3rd group is that there are 3 picks of doors (*p*) before doors are opened and each of the 3 picked doors, either the 1st, the 2nd, or the 3rd, can hide the 1 goat (*g*) that is picked in each of the 3 sequences in question. Also, the reason that there are 2 permutations after the symbol $\times$ is that there are 2 picks after doors are opened and each of the 2 doors picked, either the 1st or the 2nd, can hide the 1 pick of a goat (*g*) that is part of the sequences in question. The product $3 \times 2$ is therefore nothing but $p \times q$.

The members of the 8 groups of sequences listed in §4.7 all share the same sequence of picks once the factors have been reordered. In other words, there are only 8 *different* sequences among the 24 sequences listed in §4.7. They are as follows.

Sequence i
   $c(c-1)(c-2) \times (c-3)(c-4)$
Sequence ii
   $c(c-1)(c-2) \times (c-3)(g-o)$
 Sequence iii
   $c(c-1)g \times (c-2)(c-3)$
Sequence iv
   $c(c-1)g \times (c-2)(g-o-1)$
Sequence v
   $cg(g-1) \times (c-1)(c-2)$
Sequence vi
   $cg(g-1) \times (c-1)(g-o-2)$
Sequence vii
   $g(g-1)(g-2) \times c(c-1)$
Sequence viii
   $g(g-1)(g-2) \times c(g-o-3)$

The number of times that each of the 8 sequences is represented, namely 1, 2, 3, 6, 3, 6, 1, and 2 times respectively, may be called the *coefficient* of the 8 sequences. It has been noted above that the numbers of times in question are determined by both $p$ and $q$. It appears, therefore, that each sequence is characterized by *two* coefficients,1derived from $p$ and the other derived from $q$. It is the product of the two coefficients that constitutes the *compound* coefficient of each sequence.

The 8 compound coefficients in question can now be determined in terms of $p$ and $q$ by counting permutations of $c$ and $g$ before and after the symbol × in each of the 8 groups of sequences. The factors $p(p-1)/1 \times 2$ and $[p(p-1) \times 1 / 1 \times 2 \times p] \times q$ found in coefficients v-viii is discussed in §4.14 when the example at hand is generalized to yield an expression that applies to all possible cases.

Coefficient i
  $1 \times 1 = 1$         ($1 \times 1$ as well)
Coefficient ii
  $1 \times 2 = 2$     or    $1 \times q$
Coefficient iii
  $3 \times 1 = 3$     or    $p \times 1$
Coefficient iv
  $3 \times 2 = 6$     or    $p \times q$
Coefficient v
  $3 \times 1 = 3$     or    $\dfrac{p(p-1)}{1 \times 2} \times 1$
Coefficient vi
  $3 \times 2 = 6$     or    $\dfrac{p(p-1)}{1 \times 2} \times q$
Coefficient vii
  $1 \times 1 = 1$     or    $\dfrac{p(p-1) \times 1}{1 \times 2 \times p} \times 1 = 1 \times 1$
Coefficient viii
  $1 \times 2 = 2$     or    $\dfrac{p(p-1) \times 1}{1 \times 2 \times p} \times q = 1 \times q$

The coefficients pertaining to $p$ exhibit the sequence $1 + p + p + 1$. This 1st sequence returns to 1. The coefficients pertaining to $q$ exhibit the sequence $1 + q$. This 2nd sequence is characterized by 2 properties. First, the 2nd sequence is subordinated to, and expands, *each single* coefficient of the 1st sequence. The combined sequence is therefore $1(1+q) + p(1+q) + p(1+q) + 1(1+q)$. Second, the 2nd sequence does not return to 1. The reason for the 2nd characteristic is that the 8 sequences of picks that result in the undesired outcome of not picking a car when switching doors have been removed (see §4.9).

By uniting the coefficients i-viii with sequences i-viii, one obtains 8 products whose *sum* is the numerator of the probability that one will get at least 1 car by switching doors in the example at hand. The factor $1 \times$ is explicitly expressed for transparency.

Numerator part i: Sequence i with coefficient i
  $1 \times 1 \times c(c-1)(c-2) \times (c-3)(c-4)$
Numerator part ii: Sequence ii with coefficient ii
  $1 \times q \times c(c-1)(c-2) \times (c-3)(g-o)$
Numerator part iii: Sequence iii with coefficient iii
  $p \times 1 \times c(c-1)g \times (c-2)(c-3)$

Numerator part iv: Sequence iv with coefficient iv
$$p \times q \times c(c-1)g \times (c-2)(g-o-1)$$
Numerator part v: Sequence v with coefficient v
$$\frac{p(p-1)}{1 \times 2} \times 1 \times cg(g-1) \times (c-1)(c-2)$$
Numerator part vi: Sequence vi with coefficient vi
$$\frac{p(p-1)}{1 \times 2} \times q \times cg(g-1) \times (c-1)(g-o-2)$$
Numerator part vii: Sequence vii with coefficient vii
$$\frac{p(p-1) \times 1}{1 \times 2 \times p} \times 1 \times g(g-1)(g-2) \times c(c-1)$$
Numerator part viii: Sequence viii with coefficient viii
$$\frac{p(p-1) \times 1}{1 \times 2 \times p} \times q \times g(g-1)(g-2) \times c(g-o-3)$$

It appears that, of the 8 numerator parts listed above, i and ii share common factors, as do iii and iv, v and vi, and vii and viii. When the common factors of the 2 members of each of the 4 pairs of numerator parts are extracted and what remains is added up, one obtains 4 compound numerator parts. In the following list, the factor $1 \times$ is again retained for transparency. Furthermore, $p(p-1) \times 1 / 1 \times 2 \times p$ is the same as $1 \times$.

Compound numerator part i+ii
$$1 \times c(c-1)(c-2) \times [1 \times (c-3)(c-4) + q \times (c-3)(g-o)]$$
Compound numerator part iii+iv
$$p \times c(c-1)g \times [1 \times (c-2)(c-3) + q \times (c-2)(g-o-1)]$$
Compound numerator part v+vi
$$\frac{p(p-1)}{1 \times 2} \times cg(g-1) \times [1 \times (c-1)(c-2) + q \times (c-1)(g-o-2)]$$
Compound numerator part vii+viii
$$\frac{p(p-1) \times 1}{1 \times 2 \times p} \times g(g-1)(g-2) \times [1 \times c(c-1) + q \times c(g-o-3)]$$

The *sum* of these 4 partial compound numerators constitutes the numerator of the probability that one will get at least 1 car by switching doors. The 4 probabilities in question can be presented more compactly as follows, among others because $p(p-1) \times 1 / 1 \times 2 \times p$ is the same as 1.

i+ii:     $c(c-1)(c-2) \ [(c-3)(c-4) + q(c-3)(g-o)]$
iii+iv:   $pc(c-1)g \ [(c-2)(c-3) + q(c-2)(g-o-1)]$
v+vi:    $pcg(g-1) \ [(c-1)(c-2) + q(c-1)(g-o-2)]$
vii+viii: $g(g-1)(g-2) \ [c(c-1) + qc(g-o-3)]$

Before deriving a general expression applying to all cases from the specific example at hand, it will be useful to complete the example by computing the probability that it involves of getting at least 1 car by switching doors.

**4.11. The Probability that One Will Get at least 1 Car by Switching in the Example at**

Hand

Replacing the letters in the 4 partial compound numerators obtained at the end of §4.10 by the pertinent numbers and resolving the subtractions and the divisions yields the following partial numerators.

i+ii:     $5 \times 4 \times 3 \ (2 \times 1 + 2 \times 2 \times 5)$
iii+iv:   $3 \times 5 \times 4 \times 7 \ (3 \times 2 + 2 \times 3 \times 4)$
v+vi:     $3 \times 5 \times 7 \times 6 \ (4 \times 3 + 2 \times 4 \times 3)$
vii+viii: $7 \times 6 \times 5 \ (5 \times 4 + 2 \times 5 \times 2)$

The sum of these 4 sequences is, as it happens, exactly 45,000. This is the numerator of the probability that one will get at least 1 car by switching doors in the example at hand. The denominator is $12 \times 11 \times 10 \times 7 \times 6$ (see §4.7) or 55,440. Consequently, the probability itself is 45,000/55,440 or about 81.2%.

How does this probability compare with the probability of getting the car before switching, that is, the probability of getting at least 1 car in the 3 initial picks ($p$)? The probability of getting at least 1 car is the same as the probability of *not* picking a goat 3 times in a row in the 3 initial picks. The numerator of the probability of picking a goat 3 times in a row is $g(g-1)(g-2)$ and the denominator is $g(g-1)(g-2)$. The probability in question is therefore $7 \times 6 \times 5 / 12 \times 11 \times 10$, or 7/24, or also about 29.2%. The probability of *not* picking a goat 3 times in a row, or also of picking at least 1 car, is therefore about 70.8%.

In other words, one does somewhat increase one's chances of picking at least 1 car when switching doors, from about 70.8% to about 81.2%, by a little over 10%.

**4.12. A Key Difference between Monty Hall 1.0 and 2.0 and Monty Hall 3.0 and Higher**

What makes Monty Hall 3.0 much more interesting than Monty Hall 1.0 and 2.0 is the following. In Monty Hall 1.0 and 2.0, one *always* increases one's chances of getting 1 car by switching doors when doors are opened to reveal goats [5]. But in Monty Hall 3.0 and higher, depending on the conditions and what the desired aim is, one's chances of being successful by switching may either decrease or increase. A full study of these conditions exceeds the scope of the present paper. A complete understanding of them should make the construction of some titillating variants of the expanded Monty Hall problem possible. Some reflections follow in §6.

**4.13. First Generalization of the Numerator in the Example at Hand by Introducing Factorials**

So far, what has been obtained in regard to the example at hand is 4 compound products, the following (§4.10).

i+ii:     $c(c-1)(c-2) \ [(c-3)(c-4) + q(c-3)(g-o)]$
iii+iv:   $pc(c-1)g \ [ \ (c-2)(c-3) + q(c-2)(g-o-1)]$
v+vi:     $pcg(g-1) \ [(c-1)(c-2) + q(c-1)(g-o-2)]$
vii+viii: $g(g-1)(g-2) \ [c(c-1) + qc(g-o-3)]$

The sum of these four compound products constitutes the numerator of the probability that one will get at least 1 car by switching doors in the example at hand of the extended Monty Hall problem. How to proceed from here?

In deductive thinking, there is no need for many examples or many experiments to obtain the truth about a matter as there is in inductive thinking. The truth can be seen in, and generalized from, a single example. In deriving the general truth about the probability at hand from the example at hand, the following observation can serve as a point of departure.

The number of cars or goats decreases by 1 with each successive pick of 1 car or 1 goat. Accordingly, the sequences of products of factors listed above can be interpreted as incomplete or partial factorials or snippets of factorials. For example, the sequence of factors in the product $c(c-1)(c-2)$, in this case $5 \times 4 \times 3$, is part of the factorial $c!$, in this case $5 \times 4 \times 3 \times 2 \times 1$, or $5!$. In cases in which there are fewer cars or goats than there are picks, a factor will reduce to zero and the probability of the sequence of picks of events in question will be 0.

In a next step, the partial factorial $5 \times 4 \times 3$ can be obtained by dividing the complete factorial $5 \times 4 \times 3 \times 2 \times 1$ by the rest of the factorial, namely $2 \times 1$, or $2!$. In this case, $c(c-1)(c-2)$ equals $c!$ divided by $2!$. However, if $c$ were 6 and not 5, $c(c-1)(c-2)$ would equal $c!$ divided by $3!$. It is therefore desirable to generalize the expression of the division of a complete factorial by a partial factorial to any $c$.

In that regard, it appears that the relation between the number of the complete factorial $c!$ and the number of the partial factorial is always the same. The number of the partial factorial is always $c-3$ because the number of the picks is always 3 however many cars there are. The divisions of complete factorials by partial factorials can therefore be generalized by expressing the number of the partial factorial in its relation to the number of the complete factorial. In the case at hand, the partial factor can be expressed as $(c-3)!$ and the division of the complete factorial by the partial factorial as $c!/(c-3)!$. By this same procedure, the 4 partial numerators listed above can be converted into the following equivalents.

i+ii: $\quad \dfrac{c!}{(c-3)!} \times \quad [\dfrac{(c-3)!}{(c-5)!} + q\dfrac{(c-3)!}{(c-4)!} \times \dfrac{(g-o)!}{(g-o-1)!}]$

iii+iv: $\quad p\dfrac{c!}{(c-2)!} \times \dfrac{g!}{(g-1)!} \times \quad [\dfrac{(c-2)!}{(c-4)!} + q\dfrac{(c-2)!}{(c-3)!} \times \dfrac{(g-o-1)!}{(g-o-2)!}]$

v+vi: $\quad p\dfrac{c!}{(c-1)!} \times \dfrac{g!}{(g-2)!} \times \quad [\dfrac{(c-1)!}{(c-3)!} + q\dfrac{(c-1)!}{(c-2)!} \times \dfrac{(g-o-2)!}{(g-o-3)!}]$

vii+viii: $\quad \dfrac{g!}{(g-3)!} \times \quad [\dfrac{c!}{(c-2)!} + q\dfrac{c!}{(c-1)!} \times \dfrac{(g-o-3)!}{(g-o-4)!}]$

By being converted into $c!/(c-3)!$, an expression such as $c(c-1)(c-2)$ has been generalized to a certain degree. But it is still specific in that it only applies when $p$ is 3. In fact, all the terms in the equivalents listed above only apply when $p$ is 3 and $q$ is 2. The need is for converting the terms into expressions that apply to any $p$ and any $c$.

But before proceeding to the generalization to any $p$ and any $c$, it is necessary to detail the general structure of coefficients. The coefficients relate to how many times each of the possible sequences of picks are taken. They have already been discussed provisionally in §4.10. The need at this point is for a general treatment.

## 4.14. The Structure of Coefficients

The coefficients of the Equation pertaining to the extended Monty Hall problem exhibit the same structure as the coefficients of the power of a compound quantity that consists of two members, that is, $(a+b)^n$. The basic facts about this structure have been well-known for more than four centuries. They involve the numbers that are also found in Pascal's Arithmetical Triangle. How these numbers are obtained may be briefly reviewed below to make the present account fully self-sufficient. A particularly lucid and at the same time delightfully parsimonious presentation of the matter at hand is Euler's in his "Elements of Algebra" [6].

The *number of the coefficients* (that is, how many coefficients there are) of a compound quantity consisting of 2 members $a$ and $b$ raised to the power $n$, that is, $(a+b)^n$, equals the number of the power of the compound quantity, that is, $n$, augmented by 1, or $n+1$. The number $n+1$ is also the number of ways in which the 2 members can be arranged in regard to *how often* they are *taken*. Thus, $(a+b)^5$ yields 6 coefficients, that is, the power 5 plus 1. Accordingly, there are 6 arrangements when it comes to how often the 2 members $a$ and $b$ of the compound quantity can be taken. One can take 5 times $a$ and 0 times $b$, 4 times $a$ and 1 time $b$, 3 times $a$ and 2 times $b$, 2 times $a$ and 3 times $b$, 1 time $a$ and 4 times $b$, and 0 times $a$ and 5 times $b$. If the items are multiplied, the 6 arrangements are as follows: *aaaaa*, *aaaab*, *aaabb*, *aabbb*, *abbbb*, and *bbbbb*, which can also be written as $a^5$, $a^4b$, $a^3b^2$, $a^2b^3$, $a^1b^4$, and $b^5$. The 6 arrangements are the 6 main terms of the compound quantity. Each main term has its own coefficient.

The *coefficient numbers* (that is, what the numbers of each individual coefficient are) are determined by the number of the ways in which the 2 members of the compound quantity can be *ordered* in each of the arrangements that relate to how often they are taken. The elements can be ordered in only 1 way in *aaaaa*. Accordingly, the coefficient of $a^5$ is 1. There are 5 ways of ordering the elements in *aaaab*, namely *aaaab*, *aaaba*, *aabaa*, *abaaa*, and *baaaa*. Accordingly, the coefficient of $a^4b$ is 5. Along these same lines, the coefficients of $a^3b^2$, $a^2b^3$, $a^1b^4$, and $b^5$ can be determined to be 10, 10, 5, and 1 respectively.

In sum, $(a+b)^5$ equals $a^5 + 5a^4b + 10a^3b^2 + 10a^2b^3 + 5a^1b^4 + b^5$.

Coefficient numbers can also be obtained as follows without having to count ways of ordering elements. If all the letters are different, as in *abcde*, the number of ways in which the letters can be ordered is the factorial of the number of letters, in this case 5! If 2 letters are the same, as in *abcdd*, 5! needs to be divided by 2! Therefore, in *aaabb*, 5! needs to be divided by both 3! and 2! The result is 10. Furthermore, $5!/(3!2!)$, or $(5 \times 4 \times 3 \times 2 \times 1)/(3 \times 2 \times 1 \times 2 \times 1)$, equals $(5 \times 4 \times 3)/(1 \times 2 \times 3)$. The 6 coefficients 1, 5, 10, 10, 5, and 1 therefore equal 1, 5/1, $(5 \times 4)/(1 \times 2)$, $(5 \times 4 \times 3)/(1 \times 2 \times 3)$, $(5 \times 4 \times 3 \times 2)/(1 \times 2 \times 3 \times 4)$, and $(5 \times 4 \times 3 \times 2 \times 1)/(1 \times 2 \times 3 \times 4 \times 5)$ respectively.

The progression of the coefficients from 1st term to last term can be generalized as follows for any power $n$.

$$1, \frac{n}{1}, \frac{n(n-1)}{1 \times 2}, \frac{n(n-1)(n-2)}{1 \times 2 \times 3}, \ldots, \frac{n(n-1)(n-2)\times\ldots\times[n-(n-2)]}{1 \times 2 \times 3 \times \ldots \times (n-1)},$$

$$\frac{n(n-1)(n-2)\times\ldots\times[n-(n-1)]}{1 \times 2 \times 3 \times \ldots \times n} = 1$$

The last 2 coefficients can also be written as

$$\frac{n(n-1)(n-2)\times...\times 2}{1\times 2\times 3\times...\times(n-1)} \text{ and } \frac{n(n-1)(n-2)\times...\times 1}{1\times 2\times 3\times...\times n}.$$

The 1st coefficient is always 1 because there is only 1 way of ordering the 1st term. The last term also equals 1 for the same reason.

The coefficients involved in the extended Monty Hall problem are likewise obtained as the ways in which 2 elements can be ordered in each of the arrangements that relate to how often the 2 elements are taken. In this case, the coefficients do not equal the number of a power plus 1, but rather the number of picks of doors plus 1. The symbol *a* of the compound quantity corresponds to picking a car; the symbol *b*, to not picking a car or to picking a goat.

Each term of the compound quantity discussed above has only 1 coefficient. By contrast, each term of the probability sought in the extended Monty Hall problem has *2* coefficients if there are 2 events of picking more than 1 door and therefore 1 event of switching doors. The 1st coefficient of these 2 coefficients is derived from the number of picks in the 1st event of picking doors, that is, *p*. The 2nd coefficient is derived from the number of picks in the 2nd event of picking doors, that is, *q*. The progression of the 1st coefficient is obtained by replacing *n* by *p* in the progression listed above. The progression of the 2nd coefficient is obtained by replacing *n* by *q* in the progression listed above and leaving out the last term. The penultimate term of the progression of the coefficient *q* therefore becomes the last. It is as follows.

$$\frac{q(q-1)(q-2)\times...\times 3\times 2}{1\times 2\times 3\times...\times(q-2)(q-1)}$$

Or also as follows.

$$\frac{q(q-1)(q-2)\times...\times[q-(q-3)][q-(q-2)]}{1\times 2\times 3\times...\times(q-2)(q-1)}$$

The reason for the removal of the last term along with its coefficient is the removal of the undesired scenarios in which 0 cars are picked in the 2nd event of picking doors.

The number of the coefficients that each term has increases with, and is the same as, the number of events of picking more than 1 door. It also increases with, but is 1 less than, the number of events of switching doors.

### 4.15. The Relation between the Probability of a Sum of Partial Sequences of Picks, either Anterior or Posterior, to the Probability of the Sum of the Full Sequence of Picks

The quest involved in the Monty Hall problem and its extensions is first to establish both the probability of achieving an end by picking doors before doors are opened and the probability of achieving that same end by picking doors after doors have been opened and then to compare the two in order to determine whether, after picking doors, one improves one's chances by switching to other doors after doors have been opened.

In the example at hand, there are 32 different sequences of picking 5 doors that lead to getting at least 1 car by switching doors (§4.10) and hence 32 different numerators of the probabilities of the sequences conceived as single events. An example of a numerator is $c(c-1)(c-2) \times (c-3)(c-4)$. It pertains to the sequence in which all picks are car picks. The denominator is the same for all 32 sequences, namely $d(d-1)(d-2) \times (d-3-o)(c-4-o)$.

The probability of an individual car or goat pick conceived as a single event is expressed as a ratio of a number of available cars or goats to a number of available doors. But a sequence of picks can also be conceived as a single event. Its probability is the *product* of the probabilities that all individual picks belonging to the sequence would have if each were conceived as a single event.

Many of the probabilities of individual picks in the example at hand are conditional or dependent. A probability of an event is dependent if it is in part determined by what happens in a prior event. In other words, the probability would have been different if the earlier event had not taken place. For example, when a goat pick is the 1st goat pick of a sequence of picks, its probability, namely $g/d$, is independent. But when a goat pick is the 2nd goat pick of a sequence, the numerator of its probability will be $g-1$, one less goat being available because of what happened in the 1st goat pick. The denominator will be $d-1$ if the 2nd goat pick immediately follows the first.

Each of the 32 sequences of 5 picks in the example at hand consists of an anterior sequence of 3 picks before doors are opened and a posterior sequence of 2 picks after doors have been opened. The probability of either an anterior or a posterior sequence is the *product* of the probabilities that all individual picks belonging to the anterior or posterior sequence would have if each were conceived as a single event.

The 32 sequences of 5 picks constitute all possible cases. Furthermore, the 32 sequences are exclusive events. No 2 sequences can happen at the same time. Or, one or the other of the sequences must be the case. The *sum* of their probabilities is therefore 1 or 100%.

The 32 sequences can be collectively evaluated in search of certain properties. In the example at hand, the first 3 picks are evaluated in order to single out those sequences in which one gets at least 1 car in those 3 picks. Each sequence is an event with its own probability. Moreover, the sequences are exclusive events. The probability of all the sequences in which one gets at least 1 car in the first 3 picks is the *sum* of the probabilities of getting at least 1 car in each sequence. The 4$^{th}$ and 5$^{th}$ picks are next evaluated in order to single out those sequences in which one gets at least 1 car in those 2 picks. The probability of all the sequences of 5 picks in which this condition is met is the sum of the probabilities of the individual sequences.

But what about the probability of what happens in the 4$^{th}$ and 5$^{th}$ picks in all those sequences in which one gets at least 1 car in the first 3 picks? And what about the probability of what happens in the first 3 picks in all those sequences in which one gets at least 1 car in the 4$^{th}$ or 5$^{th}$ picks? It appears that *all possible cases* are considered in those other picks. The probability of each case will vary depending on what happens in the remaining picks of the full the sequence. But the total probability of all possible cases is 1 or 100%. It follows that, to obtain the probability of the sum of the full sequences of 5 picks that have been selected on the basis of what happens either in the anterior or in the posterior sequence of picks, one multiplies the sum of the probabilities of the anterior or the posterior sequences of picks with the total probability of either the posterior or the anterior sequences of picks, which is 1. Since multiplication by 1 does not change a number, the probability of the sum of all the *full* sequences of 5 picks that have been selected is the same as the probability of what happens either in the anterior or the posterior sequences *alone*.

Consider the example at hand, in which the aim is to get at least 1 car. Once the sequences in which one gets at least 1 car in the posterior sequences have been selected from among the 32 sequences listed in §4.8, it is possible to evaluate the total probability of all that happens in the 1$^{st}$, 2$^{nd}$, and 3$^{rd}$ picks preceding each of the selected sequences. This total probability is the sum of all the probabilities of each of the ways in which the first 3 doors can be picked. The denominator shared by all these probabilities is also the denominator of the total probability, namely $d(d-1)(d-2)$. The numerator of the total probability is the sum of the numerators of the probabilities of all 8 possible sequences of car picks and goat picks, as follows: $c(c-1)(c-2)$, $c(c-1)g$, $cg(c-1)$, $gc(c-1)$, $cg(g-1)$, $gc(g-1)$, $g(g-1)c$, and $g(g-1)(g-2)$. These 8 sequences can be brought out in front as common factors in the selected sequences of 5 picks. Thus, as the picking of doors proceeds from the 1$^{st}$ pick to the 2$^{nd}$ pick and then on to the 3$^{rd}$ pick, the numerator of the probability of what happens in the first 3 picks is the sum of the 8 combinations of 3 picks just listed and the denominator is $d(d-1)(d-2)$. In numerical terms, the sought denominator is $12(12-1)(12-2)$, or $12 \times 11 \times 10$, that is, 1320. The numerator is $5(5-1)(5-2) + 5(5-1)7 + 5 \times 7(5-1) + 7 \times 5(5-1) + 57(7-1) + 7 \times 5(7-1) + 7(7-1)5 + 7(7-1)(7-2)$, or $5 \times 4 \times 3 + 5 \times 4 \times 7 + 5 \times 7 \times 4 + 7 \times 5 \times 4 + 5 \times 7 \times 6 + 7 \times 5 \times 6 + 7 \times 6 \times 5 + 7 \times 6 \times 5$, that is 1320. The total probability is hence 1320/1320 or 1, or also 100%.

If instead the full sequences in which one gets at least 1 car in the *anterior* sequences are selected from among the 32 sequences listed in §4.8, the numerators of the probabilities of the anterior sequences of the full sequences that are being selected will be the following 7: $c(c-1)(c-2)$, $c(c-1)g$, $cg(c-1)$, $gc(c-1)$, $cg(g-1)$, and $gc(g-1)$. In other words, $g(g-1)(g-2)$ is not selected. The denominator of the same probabilities will always be the same, namely $d(d-1)(d-2)$. To compute the probability in question, a shortcut is possible (§4.11 end). The probability can be obtained by computing the probability of getting 3 goats in a row, which is the only scenario in which one does not get at least 1 car, and subtracting that probability from 1 or 100%. The probability in question is about 70.8% (§4.11 end). The total probability of the posterior sequences will be 1 because all possibilities of what can happen in the posterior picks are being considered.

In the Monty Hall problem, sums of full sequences of picks are selected first on the basis of what happens in the anterior sequences and then on the basis of what happens in the posterior sequences and the two resulting probabilities are compared. The anterior sequences will differ in the two selections. And so will the posterior sequences.

The purpose of the Monty Hall problem and its extensions is to compare the probability of sums of anterior sequences with the probability of sums of posterior sequences. Naturally, only picks belonging to anterior sequences can be considered in computing the total probability of sums of anterior sequences and the same applies in the case of posterior sequences. It is therefore not permissible, when generalizing the probabilities of the example at hand through the addition of factorials to unite into a single product probabilities of anterior car picks and probabilities of posterior car picks. Consider, for example, sequence iv in '4.10: $c(c-1)g \times (c-2)(g-o-1)$. It is possible to rearrange this sequence as $c(c-1)(c-2) \times g(g-o-1)$, bringing goat picks and car picks together. The temptation might arise to generalize $c(c-1)(c-2)$ as $c!/(c-3)!$ in an attempt to obtain a more general expression of the probability that is sought, namely of getting at least 1 car when switching doors. But the expression $c!/(c-3)!$ cannot be part of the expression of either an anterior probability or a posterior probability because it mixes elements of both.

### 4.16. Second Generalization in Terms of $p$ and $q$ of the Integers of the Example at Hand's Factorialized Numerator

The next step is to generalize the integers in the expressions at the end of §4.13 in terms of $p$ and $q$. The expressions are repeated here for ease of reference, as follows.

i+ii: $\quad \dfrac{c!}{(c-3)!} \times \quad [\dfrac{(c-3)!}{(c-5)!} + q\dfrac{(c-3)!}{(c-4)!} \times \dfrac{(g-o)!}{(g-o-1)!}]$

iii+iv: $\quad p\dfrac{c!}{(c-2)!} \times \dfrac{g!}{(g-1)!} \times [\dfrac{(c-2)!}{(c-4)!} + q\dfrac{(c-2)!}{(c-3)!} \times \dfrac{(g-o-1)!}{(g-o-2)!}]$

v+vi: $\quad p\dfrac{c!}{(c-1)!} \times \dfrac{g!}{(g-2)!} \times [\dfrac{(c-1)!}{(c-3)!} + q\dfrac{(c-1)!}{(c-2)!} \times \dfrac{(g-o-2)!}{(g-o-3)!}]$

vii+viii: $\quad \dfrac{g!}{(g-3)!} \times [\dfrac{c!}{(c-2)!} + q\dfrac{c!}{(c-1)!} \times \dfrac{(g-o-3)!}{(g-o-4)!}]$

The sum of these expressions is the probability that one will pick at least 1 car by switching doors after doors have been opened. It will be observed that, as one moves from sequence i to sequence viii, the integers pertaining to car picks *decrease* whereas the integers pertaining to goat picks *increase*. What is happening here and how does it relate to $p$ and $q$?

    At the outset of the sequences, in sequence i, the picks are all car picks. But by the end, in sequence viii, the picks are all goat picks. In each anterior or posterior sequence, there is a certain *potential* to pick cars or goats. But there is a limit to this potential. One cannot pick more cars or goats than there are picks. The *maximum* potential is therefore $p$ in anterior sequences of picks, $q$ in posterior sequences if picks, and $p + q$ in an anterior and a posterior sequence combined.

    At the outset of the sequences, in sequence i, the potential to pick cars is fully exploited. In other words, *nothing is taken or subtracted* from the potential. By contrast, everything is taken from the potential to pick goats. However, by the end, in sequence viii, it is the potential to pick goats that is fully exploited. Or nothing is taken from that potential.

    It has already been noted that the numerators and denominators of the probabilities of sequences of car or goat picks can be considered partial factorials. These partial factorials can be presented in general fashion by dividing the full factorial by the factorial whose number is the number that follows the last number of the partial factorial. For example, in the partial factorial $c(c-1)(c-2)$, the last number is $c-2$. The number that follows $c-2$ is $c-3$. The partial factorial $c(c-1)(c-2)$ can therefore be presented as the full factorial $c!$ divided by the full factorial $(c-3)!$

    At the same time, it is seen that the integer 3 is in fact $p$. After $p$ car picks, the number of available cars has decreased by $p$ and the numerator of the probability of picking a car in the next, 4[th], pick is therefore $c-3$, or generally $c-p$, because that is how many cars are still available. But this number is also the number of the full factorial by which the full factorial $c!$

must be divided to represent the sequence $c(c-1)(c-2)$ in terms of $c!$ The sequence $c(c-1)(c-2)$ can therefore be represented as $c!/(c-3)!$ By sequence viii, everything or full $p$ is taken away from the potential $p$ of picking cars. Accordingly, the numerator of the probability of picking cars may be presented as $c!/[c-(p-p)]!$, or as $c!/(c-0)!$, or also as $c!/c!$, which is the same as 1. In other words, the probability of picking cars vanishes because no cars are picked.

In the expression $c!/(c-2)!$ in sequence iii, the integer 2 is only valid when $p = 3$. In generalizing the expressions for all $p$, it appears that $2 = p-1$. Accordingly, the expression can be generalized as $c!/[c-(p-1)]!$

In the expression $c!/(c-1)!$ in sequence v, the integer is valid for all $p$, but only because sequence v is the penultimate sequence in its progression from beginning to end. As the expression $c-1$ follows $c-(p-0)$ and $c-(p-1)$ and precedes $c-(p-p)$, it can likewise be styled in terms of what is subtracted from $p$ as $c-[p-(p-1)]$.

In the following presentation, the integers in the sequences found at the end of '4.13 are interpreted in terms of $p$ and $q$. The expressions are presented as explicitly as possible for maximum transparency. They are also added up because their sum consists of the numerators of the probability of getting at least 1 car when switching doors.

(i+ii) $1 \times \dfrac{c!}{[c-(p-0)]!}$

$\times \dfrac{g!}{\{g-p-(p-0)]\}!}$

$\times [1 \times \dfrac{[c-(p-0)]!}{[c-(p-0)-(q-0)]!}$

$+ q \times \dfrac{[c-(p-0)]!}{\{c-(p-0)-[q-(q-1)]\}!}$

$\times \dfrac{\{g-o-[p-(p-0)]\}!}{\{g-o-[p-(p-0)]-[q-(q-1)]\}!} ]$

+ (iii+iv) $p \times \dfrac{c!}{[c-(p-1)]!}$

$\times \dfrac{g!}{\{g-p-(p-1)]\}!}$

$\times [1 \times \dfrac{[c-(p-1)]!}{[c-(p-1)-(q-0)]!}$

$+ q \times \dfrac{[c-(p-1)]!}{\{c-(p-1)-[q-(q-1)]\}!}$

$\times \dfrac{\{g-o-[p-(p-1)]\}!}{\{g-o-[p-(p-1)]-[q-(q-1)]\}!} ]$

+ (v+vi) $p \times \dfrac{c!}{\{c-[p-(p-1)]\}!}$

$\times \dfrac{g!}{\langle g-\{p-[p-(p-1)]\}\rangle !}$

$$\times [1 \times \frac{\{c-[p-(p-1)]\}!}{\{c-[p-(p-1)]-(q-0)\}!}$$

$$+ q \times \frac{\{c-[p-(p-1)]\}!}{\{c-[p-(p-1)]-[q-(q-1)]\}!}$$

$$\times \frac{\langle g-o-\{p-[p-(p-1)]\}\rangle!}{\langle g-o-\{p-[p-(p-1)]-[q-(q-1)]\}\rangle!}]$$

$$+ (\text{vii}+\text{viii}) \; 1 \times \frac{c!}{[c-(p-p)]!}$$

$$\times \frac{g!}{\langle g-\{p-[p-(p-1)]\}\rangle!}$$

$$\times [1 \times \frac{[c-(p-p)]!}{[c-(p-p)-(q-0)]!}$$

$$+ q \times \frac{[c-(p-p)]!}{\{c-(p-p)-[q-(q-1)]\}!}$$

$$\times \frac{\{g-o-[p-(p-p)]\}!}{\{g-o-[p-(p-p)]-[q-(q-1)]\}!}$$

Some expressions are simplified in the following equivalent. The expressions remain unambiguous while becoming somewhat less transparent.

$$\frac{c!}{(c-p)!} \times [\frac{(c-p)!}{(c-p-q)!}$$

$$+ q \frac{(c-p)!}{(c-p-[q-(q-1)])!}$$

$$\times \frac{(g-o)!}{\{g-o-[q-(q-1)]\}!}]$$

$$+ p \frac{c!}{[c-(p-1)]!} \times \frac{g!}{\{g-p-(p-1)]\}!}$$

$$\times [\frac{[c-(p-1)]!}{[c-(p-1)-q]!}$$

$$+ q \frac{[c-(p-1)]!}{\{c-(p-1)-[q-(q-1)]\}!}$$

$$\times \frac{\{g-o-[p-(p-1)]\}!}{\{g-o-[p-(p-1)]-[q-(q-1)]\}!}]$$

$$+ p \frac{c!}{\{c-[p-(p-1)]\}!} \times \frac{g!}{\langle g-\{p-[p-(p-1)]\}\rangle!}$$

$$\times [\frac{\{c-[p-(p-1)]\}!}{\{c-[p-(p-1)]-q\}!}$$

$$+ q \frac{\{c-[p-(p-1)]\}!}{\{c-[p-(p-1)]-[q-(q-1)]\}!}$$

$$\times \frac{\langle g-o-\{p-[p-(p-1)]\}\rangle!}{\langle g-o-\{p-[p-(p-1)]-[q-(q-1)]\}\rangle!}]$$

$$+ \frac{c!}{[c-(p-p)]!} \times \frac{g!}{\langle g-\{p-[p-(p-1)]\}\rangle!}$$

$$\times \left[\frac{[c-(p-p)]!}{[c-(p-p)-q]!}\right.$$

$$+ q \frac{[c-(p-p)]!}{\{c-(p-p)-[q-(q-1)]\}!}$$

$$\times \frac{\{g-o-[p-(p-p)]\}!}{\{g-o-[p-(p-p)]-[q-(q-1)]\}!}$$

This expression still reflects the values $p = 3$ and $q = 2$. There is a progression of 4 terms outside of the square brackets, 1 more than the value of $p$. Inside the square brackets, there is a progression of, not 3 terms or 1 more than $q$, but just 2 because the 3$^{rd}$ term is omitted as it concerns picking no cars in the posterior sequence of picks.

### 4.17. Generalization of the Numerator to Any *p* or *q*

In the generalization of the numerator to any $p$ or $q$, there should be $p + 1$ different coefficients in regard to the coefficient $p$ and $q$ different coefficients in regard to the coefficient $q$, that is, $q + 1$ minus 1 omitted coefficient, namely the last coefficient, which concerns picking no cars in the posterior sequence.

In the following generalized formula, there are 5 coefficients in terms of $p$, the first 4 and the last, and 5 coefficients in terms of $q$, the first 4 and the penultimate one. The sums are infinitely expandable at every instance of the expression $+ \ldots +$ . A more reduced form, still unambiguous but a little less transparent, has been anticipated in section §4.1. In case either $p$ or $q$ is equal to 3 or less, there will be fewer than 5 coefficients for either $p$ or $q$.

$$1 \times \frac{c!}{[c-(p-0)]!} \times \frac{g!}{\{g-[p-(p-0)]\}!}$$

$$\times [1 \times \frac{[c-(p-0)]!}{[c-(p-0)-(q-0)]!}$$

$$\times \frac{\{g-o-[p-(p-0)]\}!}{\{g-o-[p-(p-0)]-[q-(q-0)]\}!}$$

$$+ \frac{q}{1} \times \frac{[c-(p-0)]!}{[c-(p-0)-(q-1)]!}$$

$$\times \frac{\{g-o-[p-(p-0)]\}!}{\{g-o-[p-(p-0)]-[q-(q-1)]\}!}$$

$$+ \frac{q(q-1)}{1 \times 2} \times \frac{[c-(p-0)]!}{[c-(p-0)-(q-2)]!}$$

$$\times \frac{\{g-o-[p-(p-0)]\}!}{\{g-o-[p-(p-0)]-[q-(q-2)]\}!}$$

$$+ \frac{q(q-1)(q-2)}{1 \times 2 \times 3} \times \frac{[c-(p-0)]!}{[c-(p-0)-(q-3)]!}$$

$$\times \frac{\{g-o-[p-(p-0)]\}!}{\{g-o-[p-(p-0)]-[q-(q-3)]\}!}$$

$$+ \ldots + \frac{q(q-1)(q-2)\times\ldots\times[q-(q-3)][q-(q-2)]}{1 \times 2 \times 3 \times \ldots \times (q-2)(q-1)}$$

$$\times \frac{[c-(p-0)]!}{\{c-(p-0)-[q-(q-1)]\}!}$$

$$\times \frac{\{g-o-[p-(p-0)]\}!}{\langle g-o-[p-(p-0)]-\{q-[q-(q-1)]\}\rangle !} \Big]$$

$$+ \frac{p}{1} \times \frac{c!}{[c-(p-1)]!} \times \frac{g!}{\{g-[p-(p-1)]\}!}$$

$$\times \Big[ 1 \times \frac{[c-(p-1)]!}{[c-(p-1)-(q-1)]!}$$

$$\times \frac{\{g-o-[p-(p-1)]\}!}{\{g-o-[p-(p-1)]-[q-(q-0)]\}!}$$

$$+ \frac{q}{1} \times \frac{[c-(p-1)]!}{[c-(p-1)-(q-1)]!}$$

$$\times \frac{\{g-o-[p-(p-1)]\}!}{\{g-o-[p-(p-1)]-[q-(q-1)]\}!}$$

$$+ \frac{q(q-1)}{1 \times 2} \times \frac{[c-(p-1)]!}{[c-(p-1)-(q-2)]!}$$

$$\times \frac{\{g-o-[p-(p-1)]\}!}{\{g-o-[p-(p-1)]-[q-(q-2)]\}!}$$

$$+ \frac{q(q-1)(q-2)}{1 \times 2 \times 3} \times \frac{[c-(p-1)]!}{[c-(p-1)-(q-3)]!}$$

$$\times \frac{\{g-o-[p-(p-1)]\}!}{\{g-o-[p-(p-1)]-[q-(q-3)]\}!}$$

$$+ \ldots + \frac{q(q-1)(q-2) \times \ldots \times [q-(q-3)][q-(q-2)]}{1 \times 2 \times 3 \times \ldots \times (q-2)(q-1)}$$

$$\times \frac{[c-(p-1)]!}{\{c-(p-1)-[q-(q-1)]\}!}$$

$$\times \frac{\{g-o-[p-(p-1)]\}!}{\langle \{g-o-[p-(p-1)]-\{q-[q-(q-1)]\}\rangle !} \Big]$$

$$+ \frac{p(p-1)}{1 \times 2} \times \frac{c!}{[c-(p-2)]!} \times \frac{g!}{\{g-[p-(p-2)]\}!}$$

$$\times \Big[ 1 \times \frac{[c-(p-2)]!}{[c-(p-2)-(q-0)]!}$$

$$\times \frac{\{g-o-[p-(p-2)]\}!}{\{g-o-[p-(p-2)]-[q-(q-0)]\}!}$$

$$+ \frac{q}{1} \times \frac{[c-(p-2)]!}{[c-(p-2)-(q-1)]!}$$

$$\times \frac{\{g-o-[p-(p-2)]\}!}{\{g-o-[p-(p-2)]-[q-(q-1)]\}!}$$

$$+ \frac{q(q-1)}{1 \times 2} \times \frac{[c-(p-2)]!}{[c-(p-2)-(q-2)]!}$$

$$\times \frac{\{g-o-[p-(p-2)]\}!}{\{g-o-[p-(p-2)]-[q-(q-2)]\}!}$$

$$+ \frac{q(q-1)(q-2)}{1 \times 2 \times 3} \times \frac{[c-(p-2)]!}{[c-(p-2)-(q-3)]!}$$

$$\times \frac{\{g-o-[p-(p-2)]\}!}{\{g-o-[p-(p-2)]-[q-(q-3)]\}!}$$

$$+ \ldots + \frac{q(q-1)(q-2) \times \ldots \times [q-(q-3)][q-(q-2)]}{1 \times 2 \times 3 \times \ldots \times (q-2)(q-1)}$$

$$\times \frac{[c-(p-2)]!}{\{c-(p-2)-[q-(q-1)]\}!}$$

$$\times \frac{\{g-o-[p-(p-2)]\}!}{\langle g-o-[p-(p-2)]-\{q-[q-(q-1)]\}\rangle!} \Big]$$

$$+ \frac{p(p-1)(p-2)}{1 \times 2 \times 3} \times \frac{c!}{[c-(p-3)]!} \times \frac{g!}{\{g-[p-(p-3)]\}!}$$

$$\times [1 \times \frac{[c-(p-3)]!}{[c-(p-3)-(q-0)]!}$$

$$\times \frac{\{g-o-[p-(p-3)]\}!}{\{g-o-[p-(p-3)]-[q-(q-0)]\}!}$$

$$+ \frac{q}{1} \times \frac{[c-(p-3)]!}{[c-(p-3)-(q-1)]!}$$

$$\times \frac{\{g-o-[p-(p-3)]\}!}{\{g-o-[p-(p-3)]-[q-(q-1)]\}!}$$

$$+ \frac{q(q-1)}{1 \times 2} \times \frac{[c-(p-3)]!}{[c-(p-3)-(q-2)]!}$$

$$\times \frac{\{g-o-[p-(p-3)]\}!}{\{g-o-[p-(p-3)]-[q-(q-2)]\}!}$$

$$+ \frac{q(q-1)(q-2)}{1 \times 2 \times 3} \times \frac{[c-(p-3)]!}{[c-(p-3)-(q-3)]!}$$

$$\times \frac{\{g-o-[p-(p-3)]\}!}{\{g-o-[p-(p-3)]-[q-(q-3)]\}!}$$

$$+ \ldots + \frac{q(q-1)(q-2) \times \ldots \times [q-(q-3)][q-(q-2)]}{1 \times 2 \times 3 \times \ldots \times (q-2)(q-1)}$$

$$\times \frac{[c-(p-3)]!}{\{c-(p-3)-[q-(q-1)]\}!}$$

$$\times \frac{\{g-o-[p-(p-3)]\}!}{\langle g-o-[p-(p-3)]-\{q-[q-(q-1)]\}\rangle!} \Big]$$

$$+ \ldots + \frac{p(p-1)(p-2) \times \ldots \times [p-(p-2)][p-(p-1)]}{1 \times 2 \times 3 \times \ldots \times (p-1)p}$$

$$\times \frac{c!}{[c-(p-p)]!} \times \frac{g!}{\{g-[p-(p-p)]\}!}$$

$$\times [1 \times \frac{[c-(p-p)]!}{[c-(p-p)-(q-0)]!}$$

$$\times \frac{\{g-o-[p-(p-p)]\}!}{\{g-o-[p-(p-p)]-[q-(q-0)]\}!}$$

$$+ \frac{q}{1} \times \frac{[c-(p-p)]!}{[c-(p-p)-(q-1)]!}$$

$$\times \frac{\{g-o-[p-(p-p)]\}!}{\{g-o-[p-(p-p)]-[q-(q-1)]\}!}$$

$$+ \frac{q(q-1)}{1 \times 2} \times \frac{[c-(p-p)]!}{[c-(p-p)-(q-2)]!}$$

$$\times \frac{\{g-o-[p-(p-p)]\}!}{\{g-o-[p-(p-p)]-[q-(q-2)]\}!}$$

$$+ \frac{q(q-1)(q-2)}{1 \times 2 \times 3} \times \frac{[c-(p-p)]!}{[c-(p-p)-(q-3)]!}$$

$$\times \frac{\{g-o-[p-(p-p)]\}!}{\{g-o-[p-(p-p)]-[q-(q-3)]\}!}$$

$$+ \ldots + \frac{q(q-1)(q-2) \times \ldots \times [q-(q-3)][q-(q-2)]}{1 \times 2 \times 3 \times \ldots \times (q-2)(q-1)}$$

$$\times \frac{[c-(p-p)]!}{\{c-(p-p)-[q-(q-1)]\}!}$$

$$\times \frac{\{g-o-[p-(p-p)]\}!}{\langle g-o-[p-(p-p)]-\{q-[q-(q-1)]\}\rangle!} ]$$

**5. General Observations on Other Desired Outcomes in Monty Hall 3.0**

In the special case of Monty Hall 3.0 described in §4, the desired outcome is getting at least 1 car. But countless other outcomes may be desired. Among them are getting exactly 1 car, getting at least 2 cars, and getting exactly 2 cars, all both before and after doors are opened, as well as getting 2 cars before doors are opened and just 1 car after doors are opened. Not only the number but also the order of the picks can be specified. For example, the desired outcome might be to get at least 1 car in the last door pick, and that both before and after doors are opened. I hope to treat Monty Hall 3.0 more comprehensively elsewhere and establish the relation to the common modern probability concept of hypergeometric distribution. What follows are some general observations anticipating a more detailed treatment.

  The main observation is as follows: There is no general formula, even though certain abbreviations are possible. The basic procedure is the same for all desired outcomes in Monty Hall 3.0. First, the equation in §4.1 is expanded from just the cases in which one gets at least 1 car to all possible cases, which have a probability of 1 or 100%. The different desired outcomes are then different *selections* from the equation describing the probability of all possible cases. For fixed sequences of car picks and goat picks, coefficients need to be dropped [7].

  Suppose that the desired outcome is getting cars with every pick of a door. This is just one case of many. It is in fact a very specific case of Monty Hall 3.1. The equation in §4.1 will shrink maximally. The probability of achieving the aim at hand in the anterior picks is as follows.

$$\frac{\frac{c!}{(c-p)!}}{\frac{d!}{(d-p)!}}$$

The probability of achieving the aim in the posterior picks is as follows.

$$\frac{\frac{c!(c-p)!}{(c-p)!(c-p-q)!}}{\frac{d!(d-o)!}{(d-p)!(d-p-o-q)!}}$$

And therefore also as follows.

$$\frac{\frac{c!}{(c-p-q)!}}{\frac{d!(d-o)!}{(d-p)!(d-p-o-q)!}}$$

The factor by which one increases or decreases one's chances is then the following.

$$\frac{(c-p)!}{(c-p-q)!} \times \frac{(d-p-o-q)!}{(d-o)!}$$

### 6. Monty Hall 3.1: Some Reflections on Evaluating whether Chances of Success Increase or Decrease in Monty Hall 3.0

The original Monty Hall problem, Monty Hall 1.0, was designed as a challenge. The challenge becomes somewhat uninteresting in the expansion styled as Monty Hall 2.0 as soon as one realizes that one's chances *always increase* if doors hiding goats are opened. Nothing piques human attention more than the hope of doing better or winning or the fear of doing worse or losing, let alone the combination of the two when one is not really certain whether one will win or lose.

The quintessential uncertainty returns with Monty Hall 3.0, in which switching doors can result in either a decrease or an increase of one's chances. It is still a fact that, as in Monty Hall 1.0 and 2.0, the opening of doors always increases one's chances. Nor will one's chances decrease under those conditions if $q$ is at the same time either the same as, or larger than, $p$. However, one's chances decrease when $q$ is smaller than $p$. The key question then is whether the increase caused by opening doors is greater or smaller than the decrease caused by diminishing the number of picks from $p$ to $q$. The systematic study of the mathematical conditions that determine whether one or the other is the case in Monty Hall 3.0 may be styled provisionally as Monty Hall 3.1.

Let it suffice to present in this section examples in which the combined opening of doors and diminution of picks yields either an increase or a decrease in chances of getting a car. Let there be 6 doors and 1 car. Furthermore, let the number of picks decrease from $p$ to $q$ in that $p$

= 2 and $q$ = 1.

One's chances of getting the car in the 2 initial picks ($p$) are 11/36. That is because the chance of not picking a car in the 2 initial picks twice in succession is 5/6 × 5/6 or 25/36 and 1 − 25 / 36 is 11/36 or about 30.6%. The chance that the car is hiding behind one of the 4 remaining doors is 25/36.

If 3 of the 4 remaining doors are opened to reveal a goat, the probability of 25/36 of getting the car is compressed into the sole door that has neither been initially picked nor opened to reveal goats. One will therefore more than double one's chances of getting the car by switching doors even though one's picks are reduced from 2 to 1.

If 2 of the 4 remaining doors are opened to reveal a goat, the probability of 25/36 of getting the car is compressed into 2 doors that have neither been initially picked nor opened to reveal goats. The probability of 25/36 is distributed over those 2 doors, the chance that either door hides the car being 25/(36 × 2) or 25/72 or about 34.7%. One therefore still gains a small advantage of about 4% by switching doors.

If 1 of the 4 remaining doors is opened to reveal a goat, the probability of 25/36 of getting the car is compressed into 3 doors. The probability is therefore distributed over those 3 doors. The chance that either door hides the car is therefore 25/(36 × 3) or 25/108 or about 23.1%. In this case, one's chances of getting the car by switching decrease by between 11% and 12%.

## 7. Monty Hall 4.0: Additional Generalization to Any Number of Switches of Doors (*s*)

The generalization of the Monty Hall problem to any number of switches of doors (*s*) is styled here as Monty Hall 4.0. The following description of this generalization is limited to cases in which only 1 door is picked, as in the original Monty Hall problem (Monty Hall 1.0).

The probability of getting 1 car when switching doors any number of times is as follows, with *s* being the number of times that one switches doors, $o_1$ being the number of doors opened to reveal goats at the 1st opening of doors, $o_2$ the number of additional doors opened at the 2nd opening of doors, and so on.

$$\frac{c(d-1)(d-2-o_1)...(d-s-o_1-...-o_{s-2}-o_{s-1})}{d(d-1-o_1)(d-2-o_1-o_2)...(d-s-o_1-o_2-...-o_{s-1}-o_s)} \quad (a)$$

But before describing how this expression is obtained, it may be useful to look at the generalization at hand in a more intuitive way by means of an example.

An example is as follows. Let there be 1 car and 6 doors and therefore 5 goats. An intuitive analysis is as follows. Making a diagram may be useful in following this analysis. If one picks a door, there is a chance of 1/6 of getting the car and a chance of 5/6 that the 5 other doors are hiding the car. If 2 of those 5 other doors are then opened to reveal 2 goats, the chance of 5/6 is compressed into the 3 unopened doors of those 5 other doors. That means that each of the 3 other doors has a chance of 5/(6 × 3) or 5/18 of hiding the car. If one switches to 1 of those 3 doors, one increases one's chances of getting the car from 1/6 to 5/18. This also means that there is a chance of 10/18 or 5/18 + 5/18 that the other 2 of the 3 doors to which one could have switched hide the car. If 1 of those 2 other doors is now opened to reveal a goat in a 2[nd] round of opening doors, then the probability of 10/18 is compressed in the 1 remaining door that has been neither picked nor opened. Therefore, if one switches a 2[nd] time, now to that 1 remaining door, one doubles one's chances of getting the car from 5/18 to 10/18.

But what happens when, switching a 2nd time, one switches back to the door that was picked first? This door retains its probability of 1/6 or 3/18. In other words, after the 2 rounds of opening doors, first 2 doors and then 1 door, there are 3 doors still to be considered: (1) the door originally picked; (2) the door picked by switching; (3) the door to which one could switch by switching a 2nd time. The probabilities that these three doors hide the car are (1) 3/18 or 1/6, (2) 5/18, and (3) 10/18 respectively. This fact again illustrates the counterintuitive character of the Monty Hall problem and its extensions.

For let there be 1,000,000 doors and 1 car. At 1st pick, one has a chance of 1/1,000,000 of getting the car. There is a chance of 999,999/1,000,000 that the car is hiding behind 1 of the other doors. If 999,996 doors are now opened to reveal goats, there are 3 doors left to which one could switch. They share the probability of 999,999/1,000,000 of hiding the car and each therefore has a probability of 999,999/3,000,000. Let us assume that one switches to 1 of these 3 doors. The chances that the car is hiding behind 1 of the other 2 doors to which one does not switch are therefore 2 × 999,999/3,000,000 or 1,999,998/3,000,000. If 1 of these 3 doors is now opened to reveal 1 goat, the 1 remaining door has a chance of 1,999,998/3,000,000 of hiding the car. Thus, it may strain the imagination, but it is also undeniably true, that the 3 remaining unopened doors hold the following probabilities of hiding the car: (1) 3/3,000,000; (2) 999,999/3,000,000; (3) 1,999,998/3,000,000. Actual tests involving millions if not billions of trials, real or computer-simulated, would without any doubt confirm this fact. Similar tests of this kind have in fact been done in connection with related problems.

What happens if the door that one originally picked is opened? There are 2 possibilities. The 1st possibility is that the car is hiding behind that door. At this point, every consideration of probability instantly comes to naught because it is now 100% certain which door is hiding the car. There is no longer any probability problem because there is no longer probability but rather certainty. The 2nd possibility is that a goat is hiding behind that door. At this juncture, the situation completely changes. The door in question so far had a chance of 1/6 of hiding the car. It is now certain that it does not hide the car. The probability that it hides the car therefore drops to 0. Accordingly, the 5 other doors had so far a chance of 5/6 of hiding the car. Now it appears that these 5 doors have a chance of 100% of hiding the car. In addition, 2 doors have been opened in the 1st round of opening doors revealing goats. Consequently, the probability that the 3 other doors hide is 100%. Each of the 3 other doors therefore has a probability of 1/3 of hiding the car. At this point, we are back at the original Monty Hall problem (Monty Hall 1.0).

Now back to the generalized expression (a) above. How is it obtained? It can be obtained by generalizing the case of 2 switches of doors ($s = 2$) to any number of switches of doors. When there are 2 switches, there are 8 possible sequences of car picks and door picks, as follows: (1) *ccc*, (2) *ccg*, (3) *cgc*, (4) *gcc*, (5) *cgg*, (6) *gcg*, (7) *ggc*, and (8) *ggg*. The corresponding probabilities of the 8 sequences are as follows:

1. $\dfrac{c}{d} \times \dfrac{c-1}{d-1-o_1} \times \dfrac{c-2}{d-2-o_1-o_2}$

2. $\dfrac{c}{d} \times \dfrac{c-1}{d-1-o_1} \times \dfrac{g-o_1-o_2}{d-2-o_1-o_2}$

3. $\dfrac{c}{d} \times \dfrac{g-o_2}{d-1-o_1} \times \dfrac{c-1}{d-2-o_1-o_2}$

4. $\dfrac{g}{d} \times \dfrac{c}{d-1-o_1} \times \dfrac{c-1}{d-2-o_1-o_2}$

5. $\dfrac{c}{d} \times \dfrac{g-o_1}{d-1-o_1} \times \dfrac{g-1-o_1-o_2}{d-2-o_1-o_2}$

6. $\dfrac{g}{d} \times \dfrac{c}{d-1-o_1} \times \dfrac{g-1-o_1-o_2}{d-2-o_1-o_2}$

7. $\dfrac{g}{d} \times \dfrac{g-1-o_1}{d-1-o_1} \times \dfrac{c}{d-2-o_1-o_2}$

8. $\dfrac{g}{d} \times \dfrac{g-1-o_1}{d-1-o_1} \times \dfrac{g-2-o_1-o_2}{d-2-o_1-o_2}$

The desired outcome, getting a car after two switches of doors, is achieved in sequences (1) (3), (4), and (7). The probability of getting a car after two switches of doors is therefore the sum of the 4 probabilities (1), (3), (4), and (7).

The common denominator of this probability is as follows.

$$d(d-1-o_1)(d-2-o_1-o_2) \qquad (b)$$

And considering that 2 is the same as $s$ and 1 therefore the same as $s-1$, this expression can be rewritten as follows.

$$d[d-(s-1)-o_{s-1}](d-s-o_{s-1}-o_s) \qquad (c)$$

By extending both (b) and (c), one can derive the following expression applying to any number of switches ($s$).

$$d(d-1-o_1)(d-2-o_1-o_2)(d-3-o_1-o_2-o_3)$$
$$\times \ldots \times [d-(s-1)-o_1-o_2-o_3-\ldots-o_{s-2}-o_{s-1}] \, (d-s-o_1-o_2-o_3-\ldots-o_{s-1}-o_s) \qquad (d)$$

Expression (d) can be abbreviated as (e), without too great loss of transparency, and it has been so abbreviated in expression (a) anticipated above.

$$d(d-1-o_1)(d-2-o_1-o_2) \times \ldots \times (d-s-o_1-o_2-\ldots-o_{s-1}-o_s) \qquad (e)$$

The numerator of the probability at hand is the following sum of 4 terms found in sequences (1), (3), (4), and (7) above.

$$c(c-1)(c-2) + c(g-o_1)(c-1) + gc(c-1) + g(g-1-o_1)c$$

This sum can be rewritten as follows.

$$c(c-1)(c-2) + c(c-1)(g-o_1) + c(c-1)g + cg(g-1-o_1)$$

And therefore also in successive steps as follows.

$$c[(c-1)(c-2) + (c-1)(g-o_1) + (c-1)g + g(g-1-o_1)]$$

$$c[(c-1)(c-2+g-o_1) + g(c-1+g-1-o_1)]$$

$$c[(c-1)(c+g-2-o_1) + g(c+g-2-o_1)]$$

$$c[(c-1)(d-2-o_1) + g(d-2-o_1)]$$

$$c[(c-1+g)(d-2-o_1)]$$

$$c(c+g-1)(d-2-o_1)$$

$$c(d-1)(d-2-o_1) \tag{f}$$

And considering that 2 is the same as $s$ and 1 therefore the same as $s-1$, expression (f) can be rewritten as follows.

$$c[d-(s-1)](d-s-o_{s-1}) \tag{g}$$

By extending (f) and (g), one can derive the following expression of the numerator applying to any number of switches ($s$).

$$c(d-1)(d-2-o_1)(d-3-o_1-o_2)$$
$$\times \ldots \times [d-(s-1)-o_1-o_2-\ldots-o_{s-3}-o_{s-2}](d-s-o_1-o_2-\ldots-o_{s-2}-o_{s-1})$$

Expression (g) can be abbreviated as (h), without too great loss of transparency, and it has been so abbreviated in expression (a) anticipated above.

$$c(d-1)(d-2-o_1)(d-3-o_1-o_2) \times \ldots \times (d-s-o_1-\ldots-o_{s-2}-o_{s-1})$$

In the specific case of the above example featuring 1 car, 6 doors, and 2 switches of doors, the general expression assumes the following form.

$$\frac{c(d-1)(d-2-o_1)}{d(d-1-o_1)(d-2-o_1-o_2)}$$

Entering the relevant integers, one obtains the probability already given above.

$$\frac{1(6-1)(6-2-2)}{6(6-1-2)(6-2-2-1)} = \frac{10}{18}$$

The chances of getting a car at the first door pick is $c/d$. The factor by which one increases one's chances of getting a car after $s$ switches is therefore as follows.

$$\frac{(d-1)(d-2-o_1)\ldots(d-s-o_1-\ldots-o_{s-2}-o_{s-1})}{(d-1-o_1)(d-2-o_1-o_2)\ldots(d-s-o_1-o_2-\ldots-o_{s-1}-o_s)}$$

The factor by which one increases one's chances of getting a car from 1 switch to $s$ switches is therefore as follows.

$$\frac{(d-2-o_1)(d-3-o_1-o_2)\ldots(d-s-o_1-\ldots-o_{s-2}-o_{s-1})}{(d-2-o_1-o_2)(d-3-o_1-o_2-o_3)\ldots(d-s-o_1-o_2-\ldots-o_{s-1}-o_s)}$$

And so on.

Finally, in Monty Hall 3.0, there were 2 coefficients for each term because there were 2 events of picking doors, picking before switching and picking after switching. If doors are switched more than once, there will be more than 2 coefficients. I refrain from entering into detail at this time.

# 8. Back to Boole. By Richard D. Gill

## Summary

I comment on Leo Depuydt's recent work on applying Boole's work in probability theory to the Monty Hall problem. In particular, I compare Boole's notation and conventional modern probability notation, discuss modern computational tools, and make some comments on Boole's position that probability theory belongs to the laws of thought.

## 8.1 Introduction

George Boole's work on probability theory stands on an equal intellectual level to his work on logic, and was intended by him to be seen as an integral part thereof, yet it has largely been forgotten. Now that three half centuries have gone by and probability theory has flourished, following different routes, his work is harder than ever to read. It is impressive that Leo Depuydt has been able to find his way into Boole's way of thinking.

In this paper and its predecessor Leo tackles a number of variants of the Monty Hall problem, showing how Boole's approach leads to solutions despite ever increasing complexity. From the point of view of a present day professional mathematician, my first questions were: are the answers correct? Is Boole's probability different from present day probability?

The answers *so far* are yes: the answers are correct, and no, Boole (and with him Depuydt) is using the same probability rules as present day probabilists. I emphasize *so far* because Boole also claimed to be able to solve probability problems which modern day probabilists consider insoluble, or perhaps one could better say, ill posed. Because of this claim, influential writers of the early twentieth century such as Keynes dismissed Boole's work completely, and that hastened its progress into limbo. As Miller (2009) points out [9], however, Boole's solution was meaningful and complete, and based on adding an assumption that in absence of further information, and in particular, with no *logical* dependencies, an appropriate higher level of conditional statistical independence should be assumed. In modern day terms, Boole fitted judiciously chosen log linear models to the data, judiciously dropping higher order interactions about which there was no information anyway. This connects to modern developments in *graphical models* (also known as *Bayes nets*), another development which Boole would have appreciated, in which probability models are represented by graphs and the same graphs used as foundation for graph-theoretic based computations.

However, I do not know if Depuydt is also going to "authorize" this particular, more controversial part, of Boole's thinking.

Boole indeed saw probability theory as part of the laws of thought. His probabilities are subjective degrees of belief, their numerical values follow logically from consideration of information (known and unknown). He stood here full square in the nineteenth century tradition of Laplace, deriving probabilities from the principle of indifference, but using indifference not just to specify probabilities but also to specify probability structures.

However, so far, we are considering here problems where all probabilities are completely specified and where a frequentist (objectivist) and a Bayesian (subjectivist) approach will give the same probability values, since in either approach the symmetries of the problem force unique values of probabilities from "equally likely, by symmetry" arguments.

## 8.2 Notation

It is easiest to explain the difference between Boole's notation and modern day notation by means of a simple (mathematical) example. Consider four events which can occur, or not occur, in sequence. For instance, the results of a first pick of a door, a second pick, and so on. Let me denote the events as $A$, $B$, $C$, $D$ (capital letters early in the alphabet, according to present day conventions). The modern view of probability theory is that we may consider these events equally well as subsets of a set $\Omega$ of "elementary outcomes". The event $A$ is identified with the set of all elementary outcomes $\omega \in \Omega$ for which $A$ does indeed happen. Probabilities are assigned to subsets of $\Omega$, and set theoretic operations turn out to correspond to logical constructions involving events. For example, the event that both $A$ and $B$ occur corresponds to the outcome of the probability experiment, $\omega \in \Omega$, being both a member of the subset $A$ and the subset $B$. Thus the probability of $A$ and $B$ happening is identified with $P(A \cap B)$, where $P(\cdot)$ is a mapping from subsets of $\Omega$ to numbers between zero and one.

Subsets $A$, $B$, etc., are often called "compound events". Provided however we are careful with language, the words "elementary" and "compound" in the two contexts "elementary outcomes $\omega \in \Omega$" and "compound events $A \subseteq \Omega$", are superfluous. But it also does no harm to add them. The elementary outcomes correspond to the most fine-grained, most detailed, description of what actually happened. Compound events correspond to coarse-grained descriptions, by which many alternative "microscopic" ways according to which the same "macroscopic" phenomenon can come about are all grouped together.

Whatever probabilities are supposed to mean (whether relative frequencies in the long run of many repetitions, or whether degrees of belief as measured by fair betting odds), everyone agrees that if two events can never happen together, the probability that either occurs is equal to the sum of their probabilities; that certainty corresponds with probability one; and that all probabilities are greater than or equal to zero. Converting these minimal properties into the language of set theory, we obtain the now familiar axioms: $P(\Omega) = 1$, $P(A) \geq 0$ for all $A \subseteq \Omega$, $A \cap B = \emptyset$ implies $P(A \cup B) = P(A) + P(B)$. Finally we add as a *definition* of the *conditional probability* of $A$ given $B$, as long as $P(B) > 0$: $P(A|B) = P(A \cap B)/P(B)$.

From these minimal properties one can derive the following chain rule:

$$P(A \text{ and } B \text{ and } C \text{ and } D) \;=\; P(A)P(B|A)P(C|A \text{ and } B)P(D|A \text{ and } B \text{ and } C). \quad (1)$$

However, the alert reader will have noticed that I am mixing the language of logic and the language of elementary set theory in this equation, and I do that deliberately, in order to point out an important ambiguity in the translation from logic to set theory.

The interpretation of the left hand side is obvious: I could have written (should have written!), of course, $P(A \cap B \cap C \cap D)$. The right hand side certainly makes sense, and indeed the statement is true, if I do the corresponding substitutions on that side; for instance, the last term should be $P(D|A \cap B \cap C)$. However there is an alternative substitution, more clumsily expressed in set theoretic language, but equally meaningful from the point of view of natural language. The correctness of this alternative interpretation is actually a theorem.

Let me explain. Suppose I start with a probability measure $P$. Next I pick some event $B$ with positive probability, and compute new probabilities $P^B(A) = P(A|B)$ for *every* event $A$.

**Theorem 1**: the conditional probability measure $P^A$ also satisfies the axioms of probability theory;

**Theorem 2**: (principle of repeated conditioning): $(P^A)^B = P^{A \cap B}$.

This is not just empty formalism, it tells us something very important: conditioning in turn on any number of events gives end results which do not depend on the order in which we take them, and is also not changed by grouping them into a smaller number of events by using the rule $P(A|B \text{ and } C) = P(A|B \cap C)$. It shows us that the transition between the language of logic and the language of sets is very smooth indeed.

Boole has no use for the language of sets. It was not even yet invented: his supporter and contemporary John Venn was one of those who pioneered its use; indeed, its use in probability theory. For Boole, the language of logic does fine *both* for events *and* for probabilities of events. Defining the event $E$ as "$A$ and $B$ and $C$ and $D$", Boole writes the definition of the *event E* as

$$e = abcd \quad \text{(logical relation)}, \qquad (2)$$

and then rewrites equation (1), a relation between probabilities and conditional probabilities, with the very same sequence of symbols:

$$e = abcd \quad \text{(numerical relation between probabilities)}. \qquad (3)$$

Even though equation (3) is to be interpreted numerically as a relation between probabilities, the rules of algebra have to be handled with very great care. The exact sequence of probabilities *abcd* corresponds to a specified sequence of events $A$, $B$, $C$, $D$ and there is a logic to this sequence: typically this will be their *temporal* ordering. The value assigned to the numerical variable $c$, for instance, depends on the context, on the presence in front of *events a* and *b*. Event $D$ might be certain in some context, impossible in other. The preceding events $A$, $B$ and $C$ could switch the *probability d* to 1, or to 0. This is what Depuydt calls the digital nature of probabilities.

## 8.3 Computations

One of the fruits of the digital revolution has been statistical computing and computer algebra. Looking at the huge tables of probabilities in Sections 4.1 and 4.17, the reader may worry that perhaps some typesetting error has corrupted one of the formulae. If the reader actually wanted to use those formulae to do numerical computations, he or she might want a computer do those computations. But then the typeset symbols have to be translated into lines of computer code, which is another opportunity for errors to creep in.

I have verified that it is in principle possible to reproduce these tables using computer algebra. Let the computer do the painstaking, repetitive task of applying simple rules of transformations of formulas! Let the computer typeset the pages in the journal, let the computer also generate computer code for implementation to specific cases! Then the reader need only check the programs or scripts: do they implement Boole's logic of probability? There are two levels involved here. The *problem* should be described in a high level formal language which translates line by line Depuydt's verbal descriptions of what he is doing into a language which the computer algebra system knows. Anyone who understands the language can verify that it is "the same thing". The implementation of the computer algebra system must be checked by specialists, though users who use it day by day for a myriad of tasks also provide external consistency checks whenever the answer can be got by different means.

I would like especially to draw the reader's attention to two powerful tools, both of them completely free (both in the sense of "free beer" and in the sense of "free speech"): the statistical language R (http://R-project.org) and the computer algebra system Sage (http://sagemath.org). The freedom as in free speech is the fact that the computer code of both R and Sage itself are publicly available, and anyone is allowed not only to look at it but also to modify it, repackage it, and even to sell it, as long as their modifications preserve the same freedoms.

*Sage* allows one to to instruct the computer to perform algebraic formula manipulations according to specified rules. Boole would have appreciated that. Unlike commercial tools like Mathematica, the algorithms which it uses are public; the scientist can check them, even replace them by new algorithms of their own.

*R* is a statistical computing tool. One thing which is extremely easy with R is to run a computer simulation of millions of repetitions of a particular generalized Monty Hall problem, count outcomes of different kinds, in order to statistically estimate the probabilities which can in principle be computed algebraically.

Both these systems are widely used in academia, in teaching, in industry; they have huge followings and because of their open nature, additions have been written by users from all kinds of application fields which anyone else can also freely use. The user communities with their internet fora and mailing lists and so on, allows both the new user and the expert to get advice from fellow users all over the world, often extremely rapidly and effectively. R can even be used from Sage – one of the design philosophies of Sage is to use existing tools, so as not to waste time re-engineering wheels. This has certainly paid off, since in a short time Sage has become extremely powerful and flexible. Like natural languages, culture, and like mathematics itself, these systems evolve through highly effective "crowd-sourcing".

## 8.4 Alternative approach

Depuydt goes back to first principles and determines the probabilities of all possible *elementary outcomes* of his Monty Hall games: any particular sequence of picks of doors. Now, it is possible to group some of the picks together, producing a coarser level of description, but one in which (a) the components of the coarser description correspond to familiar probability models, and (b) the coarser description is fine enough to still allow specification of the *compound events* of interest.

In Monty Hall 3.0, such a coarser description is possible at the level of *phases*. Recall that in this game, $c$ doors hide cars, $g$ doors hide goats, $d = c + g$ is the total number of doors. The player first picks $p$ doors. The host then opens $q$ doors, revealing goats. The player may now switch to another $r$ doors.

The hosts' possibilities are delimited by how many cars are hidden by the player's first $p$ picks. Call this number $x$. We can now write down the joint probability of $x$ cars being behind the player's first $p$ picks, and $y$ cars being behind the player's second $r$ picks, as follows. Both phases correspond to a traditional "sampling without replacement" situation, picking balls from vases, where the composition of the vase at phase two is determined by the outcome of ball-picking in phase one.

Suppose a vase contains $R$ red balls and $B$ blue balls, let $N = R+B$ be the total number of balls in the vase. Suppose $n$ balls are picked at random from vase, without replacement, and completely at random. Define the binomial coefficient $C^n_x = n!/x!(n-x)!$, the number of ways to choose $x$ objects from a collection of $n$. In spoken mathematics, one says "$n$ choose $x$" instead of "$C$ superscript $n$ subscript $x$". Let $r$ be the number of red balls in the sample of $n$, and define $b = n - r$ to be the number of blue balls. It turns out that the probability to find exactly $r$ red balls is $h(r; n, R, N) = C^R_r C^B_b / C^N_n$. The fact that these so-called *hypergeometric* probabilities must add up to one as one adds over all possible values of $r$ is called the Chu-Vandermonde identity in combinatorics, going back to Chu Shi-Chieh, 1303, and Alexandre-Théophile Vandermonde, 1772. One can say that Depuydt has derived a "two-level" generalization of this identity from first principles, following Boole's methodology.

Now if among the first $p$ doors chosen by the player exactly $x$ doors hide cars, then at the second stage, when there are $d-p-q$ doors left from which the player may choose $r$ doors, a further $q$ doors already having been opened revealing goats, exactly $c-x$ of those doors hide cars, and $g-(p-x)-q$ hide goats. This tells us that the probability that the player's first $p$ picks hide $x$ cars and his second $r$ picks hide $y$ cars is $h(x; p, c, d)h(y; r, c-x, d-p-q)$.

This gives an alternative way to check the results of this paper.

## 9. Empirical Definition of Mathematics, in Boole's Footsteps, as a Cognitive Event on the Deepest Level

### 9.1. Where Is Mathematics?

The question that is at the center of the present section is as follows: What is mathematics? The answer to this question is of considerable interest. No endeavor of the human intellect has been more successful. Evidently, the question has occupied many, many minds over the centuries. The literature on the subject is massive. But even the most cursory review of what has been done readily reveals that the question can hardly be considered answered. There has been no lack of attempts to provide an answer. However, the proposed answers seem often irreconcilable and can even be diametrically opposed.

In order to define mathematics, one needs to be able to observe it. A second question therefore presents itself, as follows: Where is mathematics? In other words, where can one find mathematics so that one can take a look it and analyze it in order to determine what it really is?

It appears that the answer to the seemingly simple question as to *where* mathematics is has perhaps been the greatest point of controversy in the discussion of *what* mathematics is. There have been basically two diametrically opposed answers to the question where mathematics is. Some believe that mathematics is something inside the head. Others believe that it is something outside the head. Whereas many believe with Kurt Gʻdel that numbers exist independently of the human mind, many others like L.E.J. Brouwer are convinced that numbers are a creation of the human mind. Could both be right at the same time?

The position that I will adhere to is that mathematics can only be *empirically observed* as something that is *inside* the head. This position in no way involves a denial of the notion that mathematics is something outside the head. Clearly, when applied to reality outside the head, mathematics works. Some may interpret this as proof that mathematics is also something outside the head. Then again, the totality of human experience of reality outside the head is how the brain perceives and processes this reality through the senses inside the head. This perception is itself 100% brain activity. Therefore, the analysis of the human experience ought to consist in the final resort, on the deepest level, of the analysis of brain activity. And one component of the brain activity that constitutes the human experience is mathematics. In that regard, the question as to whether mathematics is also something outside the brain is to some extent moot because mathematics cannot be empirically observed in that capacity anyhow, so there is hardly anything to be said about that capacity.

To some extent, Gʻdel's position and Brouwer's position are not in opposition. There is nothing that contradicts the notion that numbers are something that is both something outside the head and something inside the head. Naturally, numbers would inhabit different mediums inside the head and outside the head, physical reality and brain mass respectively. But when it comes to empirical observation, numbers can only be observed and therefore also analyzed as an activity of the brain or an event happening to the brain. Mathematics is something that the brain does.

The design of what follows is to resume a line of inquiry that has been abandoned about a century and a half ago. It appears that this line of inquiry is based on the assumption that mathematics is something the brain does.

## 9.2. Resuming an Abandoned Line of Inquiry

The aim of what follows is to pursue a line of inquiry that was initiated about a century and a half ago but fairly soon completely abandoned and ever since entirely disregarded. This line of inquiry is, I believe, worthy of being resumed. It appears to me that it can lead to a final definition of mathematics and its foundations. The initiator of the line of inquiry in question was George Boole, first in his *The Mathematical Analysis of Logic* (1847) [9], but then above all in his *An Investigation of the Laws of Thought* (1854), which may be regarded as the Magna Charta of the digital age [10]. The principal follower of Boole was John Venn in his *Symbolic Logic* (second edition, 1894) [11]. Whitehead notes that Venn gave "thorough consistency to Boole's ideas and notation, with the slightest possible change" [12] and, more recently, Styazkhin observed that Venn "revealed the essence of the secret of success of Boole's procedures" [13].

The Digital Age owes an extraordinary debt to Boole and to the digital mathematics that he created. Digital mathematics is a type of mathematics that is distinct from the more familiar type of mathematics, quantitative mathematics (to which Boole also made significant contributions, for example by his work on differential Equations). But digital mathematics is in the end just as mathematical as quantitative mathematics. Clearly, a line of inquiry initiated by Boole has proved to be successful. Little did Boole know to which uses his digital mathematics would be put when he wrote in 1847, "It would be premature to speak of the value which this method may possess as an instrument of scientific investigation" [14].

## 9.3. Probability Theory as an Ulterior Aim of Said Line of Inquiry

It is not clear to which extent Boole, when initiating the line of inquiry that ultimately spawned the Digital Age, had something like computer science in mind as an ulterior aim. In fact, in his *Laws of Thought*, digital mathematics is clearly subordinated to an ulterior aim of an entirely different kind, namely making classical probability theory complete.

It is not entirely certain whether Boole had this relation of subordination to probability theory in mind as soon as he began working on digital mathematics. There is no mention of probability theory in his *The Mathematical Analysis of Logic* of 1847, in which he first established his digital mathematics. But in *Laws of Thought*, digital mathematics is clearly styled as serving the aims of probability theory, as appears from the second part of the book's long title, *(An Investigation of the Laws of Thought,) on Which Are Founded the Mathematical Theories of Logic and Probabilities*.

A great irony relating to Boole's legacy is that his work on probability theory has been, with one or two exceptions [15], completely disregarded, almost entirely bypassed by the field. In the planned article mentioned at the end of §1, I intend to confirm that Boole's probability theory does what it claims to do, make classical probability theory complete, and how it does so.

## 9.4. Rational Thought and Language as an Ulterior Aim of Said Line of Inquiry

But there seems to be more to the ulterior aims of Boole's digital mathematics than statements about probability theory. Boole's *Laws of Thought* and many of his other works on logic and probability, both published and unpublished, are replete with references to the nature of human thought in as far as thought is rational. The question as to whether he aimed to determine what is going on in one's head when one thinks rationally is investigated below.

In any event, like Boole's ideas on probability theory, this component too of his line of inquiry appears to have fallen by the wayside. Whereas the forthcoming article mentioned at the end of §1 is an attempt to validate Boole's line of inquiry in relation to probability theory, what follows is an attempt to resume and extend this same line of inquiry as it relates to the deepest foundations of rational human thought and mathematics.

## 9.5. Is Boole's Work on Logic and Probability Mathematical or Cognitive in Nature?

### 9.5.1. Modern Perception of Boole's Work on Logic as Strictly Mathematical
When one reads Boole's writings on logic and probability, the following question easily arises: Is Boole doing mathematics or is he trying to determine how people think rationally? In other words, is he describing the mathematical structure of reality or is he trying to tell us what is going on in people's heads when they think rationally? Boole's contributions, to the extent that they have proved lasting, are now universally perceived as belonging to the realm of mathematics. Boolean algebra is after all ubiquitous. Bertrand Russell even accused Boole of giving his 1854 book the wrong title. He believed that Boole was "mistaken in supposing that he was dealing with the laws of thought," because "the question how people actually think was quite irrelevant to him" [16]. Taking into consideration how people think while practicing mathematics is sometimes called psychologism, which some seem to regard as a bad word.

### 9.5.2. Statements to the Contrary in Boole's Writings
There are abundant indications in Boole's work that leave no doubt that how people think, at least as far as rational thought and language is concerned, was very much on his mind. In the Preface to the earlier *Mathematical Analysis of Logic* (1847), he states that he is not concerned with "quantity," but with "facts of another order which have their abode in the constitution of the Mind" [17]. In the first statement following the Preface to the later *Laws of Thought* (1854), he announces [18]:

> The design of the following treatise is to investigate the fundamental laws of those operations of the mind by which reasoning is performed.

How can such statements, when taken at face value, not pertain to what is going on inside the heads of people—notwithstanding attempts to soften their impact, perhaps to protect Boole from the charge of psychologism? [19]

Two possible reasons for resisting the notion that Boole could have been aiming to establish how the brain works are as follows.

First, mixing Boole's mathematical results with reflections on the nature of thought might be seen as affecting the purity and objectivity of the former.

Second, at the present time, it remains still basically unknown—let alone that it was in Boole's time—how the brain produces rational thought and language in biochemical terms, that is, which activities of neurons and synapses are responsible. So how could anyone have anything to say about how the biological brain reasons?

The first objection is addressed below. In regard to the second objection, I have noted elsewhere that, as one brain communicates with another through thought and language, all communications need to travel by air from the mouth of a speaker to the ear of a hearer or by light from the written page to a reader's eyes. There can be no doubt that everything that is essential to the structure of rational thought and language must be conveyed in sound waves or light beams that travel from mouth to ear or from page to eye. In that sense, the structure of rational thought and language is empirically accessible. The same structure ought to be present inside the brain, even if inhabiting a different medium.

### 9.5.3. Boole's Own Perception of His Work on Logic as Mathematical

While there can be no doubt that how the brain thinks is somehow a prominent concern in Boole's writings, there are also plenty of statements in his writings that leave no doubt that he is firmly convinced that what he is doing when he is studying logic and probability is mathematics. He states, for example, that "the ultimate laws of Logic are mathematical in their form" [20].

As one tries to assess what exactly it is that Boole is trying to do, the impression gradually imposes itself and becomes inescapable that he is writing both about how the mind thinks and about mathematics. There are just too many categorical statements in his work that positively point to both. At this juncture, there is the possibility of assuming that there is something deeply confusing and contradictory in Boole's work. One might seek to resolve the possible contradiction by discarding either the cognitive facet or the mathematical facet of Boole's work as invalid. In choosing to reject either of the two, the easier choice would seem to be the cognitive facet. The mathematical facet has more than proved itself by applications in modern computer science.

Then again, it is difficult to overlook the many passages that concern how the mind thinks. Consider his analysis of the syllogism, which does not supplant Aristotle's analysis but rather completes it. It seems easy for all to agree that we must think according to the rules of the syllogism if we are to reason correctly. And more generally, it is easy to convince oneself that what Boole says about how the mind thinks rings true. There just seems to be more to Boole's writings than just mathematics.

### 9.5.4. Could Boole's Work on Logic Be Both Mathematical and Cognitive?
The question arises: Could Boole have been doing both at the same time, producing mathematics and describing mental faculties? The following statement by Boole clearly indicates that his approach is at the same time mathematical and cognitive. What he sets out to discover is the mathematical structure of rational cognition [21]:

> The laws we have to examine are the laws of one of the most important of our mental faculties. The mathematics we have to construct are the mathematics of the human intellect.

The present discussion has reached a critical juncture. It needs to be decided whether the cognitive facet of Boole's line of inquiry should be pursued or dropped altogether. The validity of Boole's digital mathematics says something about the overall soundness of his thinking. It can serve as an argument in favor of resuming the cognitive facet of the same general line of inquiry.

In resuming the cognitive facet, the concept described by Boole as the "mathematics of the human intellect" cited above will serve as a point of departure. What can possibly be meant by this concept? It would seem that it places mathematics somehow inside the human intellect. The way in which the concept will be interpreted in what follows is that mathematics is in essence a property and an activity of the brain. Mathematics is best defined as something that the brain does. In a planned article, I hope to show that probability theory is a good illustration of this definition. It seems otherwise quite tempting to interpret mathematics as exactly the opposite, namely as a property of reality outside the brain. As it happens, that very notion will also be assigned a place in the definition of mathematics as something that the brain does. Meanwhile, the principal consideration that leads to the definition of mathematics as an activity of the brain is presented in the next section.

## 9.6. Mathematics as an Activity of the Brain

The brain is evidently the most complex structure in the universe. It consists of billions of neurons and trillions of synapses. Still, it seems just as evident that the brain is a biological mass that is limited in size. There is only so much of it and no more. The following working hypothesis therefore seems to impose itself. The time will come when it will be possible to record everything that the brain does as it happens, presumably with the aid of supercomputers or the like. The opposite of this hypothesis is that a certain part of the brain will be forever inaccessible. But what could such an inaccessible part consist of? If everything in the brain is atoms and molecules and the like, then no activity in the brain should in the end avoid detection, one would think.

Another basic assumption is that the totality of human existence as we know it consists of how the brain perceives reality outside itself through the senses. There are many more senses than the classic five, including sensing the effects of the instincts with which the brain comes equipped at birth. In addition, perceptions received through the senses can be recombined in certain ways inside the brain. Dreams are one type of recombination. The opposite of this assumption is that there is something more to reality than what is perceived through the senses. It is difficult to see what that something more could be. Religion makes certain assumptions about that something more. But then, it is impossible to make everyone agree on what that something more is and the assumptions of religion are beyond scientific verification anyhow.

Once it is possible to record everything the brain does in its entirety, part of what is recorded will be the brain's knowledge and practice of mathematics. It should be possible to observe exactly what the brain does when it engages in mathematics and how it starts up mathematical knowledge. The key question arises: Is there more to mathematics than recorded brain activity?

**9.7. The Brain as the Final Frontier: Towards a New Empiricism**

If the totality of the human experience consists of how the brain engages what is outside itself, then nothing that does not have some kind of imprint in the brain can mean anything to the brain. In assessing what is outside itself, the brain only has itself, as it were, to sort things out. And by itself is meant a complex and very large but ultimately limited and fully definable amount of activity of neurons and synapses and the like.

At first sight, it would seem as if mathematics is a property of reality in which the brain occasionally participates. Mathematics seems like a sacred code inscribed in the book of nature. But all that the brain can ultimately know about this code is the details of its own participation. And the details of this participation consist one hundred percent of brain activity. Therefore, if one truly wants to understand what mathematics is, then all one has as an object of study is the participation itself as brain activity. It is understandable that there may be a desire for more than just that. But the brain can hardly step outside itself, as it were. It is fully limited to its own activity and powers, and to the study of this activity and these powers in a search for understanding. The brain activity does not only include mathematical knowledge and reasoning, but also the act of perception in the form of signals reaching the brain from outside through the senses. Needless to say, once it is possible to observe all this brain activity, it will also be possible for this very act of observing brain activity to be itself observed, including by the person whose brain activity is being observed. It is a bit like a snake biting its own tail.

But what about the ever attractive notion that mathematics is a property of nature outside the brain? Nothing is more tempting than to subscribe to this assumption. In fact, I believe that there is nothing wrong with *assuming* that reality exhibits a structure that may be called mathematical and that this structure is somehow the origin of a certain type of brain activity that may be described as knowing and doing mathematics. It is an impossible to avoid assumption under which everyone effectively operates. One way of looking at the matter is as follows. It is not because there is no final verification of this assumption that the assumption should be rejected. The assumption receives abundant support from the fact that mathematics works. When mathematical knowledge is acquired and this knowledge is then returned to reality outside the brain by being applied correctly, as in building a bridge, the application will typically work, that is, the bridge will not collapse. But ultimately, mathematics can only be observed to the extent that it can be seen at work in the brain. Reality is experienced entirely in terms of how the brain engages what is outside itself through the senses. The scientific observation and analysis of this experience therefore ultimately needs to be the observation of the brain. And that also applies to mathematics as one type of reality. Anything else is beyond human knowledge. It is not possible to look behind the curtain, as it were, to establish why the brain is the way it is. Along these lines, Boole's writes in somewhat Latinate English, "It may, perhaps, be permitted to the mind to attain a knowledge of the laws to which it is itself subject, without its being also given to it to understand their ground and origin, or even, except in a very limited degree, to comprehend their fitness for their end, as compared with other and conceivable systems of law" [22].

Because the assumption that the structure of reality outside the brain is mathematical is just an assumption, it is not possible to probe the deeper roots of this presumed structure. There is of course nothing that prevents anyone from engaging in speculation to any degree. It is likewise possible to speculate without restrictions about other possible types of realities in which other possible types of mathematics apply.

Knowledge is ultimately a process of assimilation in which the brain assimilates to reality outside itself. For example, to find one's way through the streets of a city without consulting a street map, the brain needs to acquire something of the structure of the layout of the city's streets and in that sense become a little like that layout. But it is reasonable to assume that, in the process of assimilation, there needs to be something to assimilate to. Therefore, if part of the assimilation is mathematical, there is presumably something mathematical in reality outside the brain to which the brain assimilates.

The fact that the knowledge of mathematics is stored in the books of a mathematics library may also seem to suggest that mathematics is something outside the head and hence first and foremost a property of nature, with its reflection inside the head being somehow secondary. However, the books in question are nothing more than paper and ink until an active brain reads and studies them. In that sense, a tree does not fall in the forest if there is no one there to hear it. The mathematics in a book is not mathematics if it is not actively engaged by a thinking brain.

### 9.8. Conclusion

It is possible to reconcile as complimentary the view adhered to by someone like Brouwer that mathematics is something inside the head and the view adhered to by someone like Gödel that mathematics is something outside the head. In other words, the two views do not contradict one another. However, that mathematics is something outside the head is only an assumption. But it is an assumption that is hard to deny. So to some degree Gödel's view can be recognized. Still, it is only as something inside the head that mathematics can be truly observed and therefore become the subject of empirical inquiry once the secrets of the brain are unlocked. In that regard, the cognitive approach is the only one that offers a systematic path of scientific investigation. I hope to apply the cognitive approach in planned papers, beginning with the branch of mathematics called probability theory. It will be useful to formulate the foundations of probability theory fully in cognitive fashion.

### 10. Acknowledgments

When the present article was essentially complete, a fortunate set of circumstances brought the author in contact with Dr. Richard D. Gill, Professor of Mathematical Statistics at the Mathematisch Instituut of Leiden University in the Netherlands, who has contributed a number of studies to the analysis of the Monty Hall problem. An interesting exchange of ideas ensued about all sorts of facets of the expanded Monty Hall problem and about the contents of the present article. I personally profited much from this exchange. One result of the exchange was the decision to include, for the benefit of a somewhat more interdisciplinary audience, an appendix by Professor Gill (see §8). The design of this appendix is to provide additional context by building a bridge to modern probability theory in its conventional notation and by pointing to the benefits of certain interesting and relevant tools of computation now available on the Internet. A more detailed and in-depth description of the common concept of the hypergeometric distribution in its relation to the contents of the present article remains desirable and will need to be postponed to future papers. The present collaboration is meant as a first step conceived and executed on short notice, an exploratory effort that probes what is possible in terms of interdisciplinary projects spanning both the humanities and the sciences. I am grateful to Professor Gill for his willingness to make this much appreciated contribution.

As regards the interdisciplinary nature of the larger project of whose mathematical branch this article is part, one ulterior aim is to promote the perfect unity of notation of Boole's algebra and the complete unity of human intelligence that it suggests, in that the notation can be applied to the following multiple facets of human intelligence, each illustrated here by one expression.

(1) "The sun shines." (*Language, Level of the Things*)

(2) "When the sun shines, I go to the beach." (*Language, Level of the Events*)

(3) "Humans are mortal. Socrates is mortal. Therefore, Socrates is a human being." (*Logic*, but not just with three statements, as in the present example, but with any number of statements)

(4) A quadratic equation. (*Quantitive mathematics*)

(5) The Monty Hall problem. (*Digital mathematics*, in addition to quantitative mathematics)

Finally, I also thank Dr. Michael R. Powers, now Professor of Risk and Insurance Mathematics at Tsinghua University in Beijing, China, for his continued interest in the subject matter of this article and for fruitful discussions in this connection about the essence of probability in both its strictly mathematical and its more subjective interpretations.

## 11. References

[1] L. Depuydt, "The Monty Hall Problem and Beyond: Digital-Mathematical and Cognitive Analysis in Boole's Algebra, Including an Extension and Generalization to Related Cases," *Advances in Pure Mathematics*, Vol. 1, No. 4, July 2011, pp. 136-154. For the history of the problem and its context, I refer again to J. Rosenhouse's recent book, "The Monty Hall Problem," Oxford University Press, Oxford and New York, 2009. More recently, the slim volume by R. Deaves, *The Monty Hall Problem: Beyond Closed Doors*, 2006 (published by www.lulu.com) serves as an additional illustration of the widespread interest in the Monty Hall problem. A principal catalyst in the rise of interest in the problem was an answer by Marilyn vos Savant, who according to some sources was supposed to be the holder of the world's highest IQ (Rosenhouse, "The Monty Hall Problem," p. 29), in a column in the magazine *Parade* of September 9, 1990 to a question by a reader about the Monty Hall problem. The answer provoked a flurry of activity (for a detailed account of this activity, see Rosenhouse, "The Monty Hall Problem," p. 23-31).

[2] H.H. Goldstine, "The Computer from Pascal to von Neumann," Princeton University Press, Princeton, 1972, p. 37.

[3] L. Depuydt, "The Monty Hall Problem and Beyond: Digital-Mathematical and Cognitive Analysis in Boole's Algebra, Including an Extension and Generalization to Related Cases" [see note 1], p. 148.

[4] A.W.F. Edwards, "Pascal's Arithmetical Triangle: The Story of a Mathematical Idea," Johns Hopkins University Press, Baltimore, 2002, p. xiii.

[5] L. Depuydt, "The Monty Hall Problem and Beyond: Digital-Mathematical and Cognitive Analysis in Boole's Algebra, Including an Extension and Generalization to Related Cases" [see note 1], p. 145.

[6] L. Euler, "Elements of Algebra," Springer Verlag, New York, Berlin, Heidelberg, and Tokyo, 1984 (reprint of the English translation of 1840 by J. Hewlett of, [1], the French translation by J. Bernoulli (III) of the original German edition of 1770 and of, [2], the French additions by J.-L. Lagrange, with a preface by C. Truesdell that earlier appeared in H.E. Pagliaro [ed.], "Irrationalism in the Eighteenth Century," The Press of Case Western Reserve University, Cleveland and London, 1972, pp. 51-95), pp. 110-120.

[7] "Si l'on voulait établir une succession determine, il faudrait supprimer le coefficient," writes S.F. Lacroix, "Traité élémentaire du calcul des probabilités," Fourth edition, Mallet-Bachelier, Paris, 1864, p. 30.

[8] The procedure resembles the operations described in L. Depuydt, "The Monty Hall Problem and Beyond: Digital-Mathematical and Cognitive Analysis in Boole's Algebra, Including an Extension and Generalization to Related Cases" [see note 1], p. 150, §4.7.

[9] G. Boole, "The Mathematical Analysis of Logic, Being an Essay towards a Calculus of Deductive Reasoning," Macmillan, Barclay, & Macmillan, Cambridge and George Bell, London, 1847. I have used the reprint in, and pagination of, G. Boole, "Studies in Logic and Probability," Dover Publications, Mineola, New York, pp. 45-124 (which is itself a reprint of G. Boole, "Studies in Logic and Probability," Watts & Co., London, 1952). It should be noted that Boole's "0" is electrical engineering's "1" and *vice versa*, Boole's "⋂" (AND) is electrical engineering's "+" (AND) and *vice versa*, and Boole's "+" (OR) is electrical engineering's "⋂," facts that I have failed to appreciate in the introduction to my "The Other Mathematics: Language and Logic in Egyptian and in General," Gorgias Press, Piscataway, New Jersey, 2008, even if this oversight does not affect the arguments presented in this work. It is difficult to find any published observations anywhere pointing explicitly to these fundamental facts.

[10] G. Boole, "An Investigation of the Laws of Thought, on Which Are Founded the Mathematical Theories of Logic and Probabilities," Walton and Maberly, London, 1854. I have used the reprint of 1958 by Dover Publications, New York.

[11] J. Venn, "Symbolic Logic," second edition, Macmillan and Co., London and New York, 1894.

[12] A.N. Whitehead, "A Treatise on Universal Algebra with Applications," Cambridge, 1897, p. 11.

[13] N.I. Styazhkin, "History of Mathematical Logic from Leibniz to Peano," Cambridge, Mass., 1969, p. 214. The original Russian version of this work appeared in 1964.

[14] G. Boole, "The Mathematical Analysis of Logic" [see note 6], p. 53.

[15] Most notably Th. Hailperin, "Boole's Logic and Probability," Second edition, Revised and enlarged, Studies in Logic and the Foundations of Mathematics 85, North-Holland Publishing Company, Amsterdam, New York, Oxford, and Tokyo, 1986.

[16] B. Russell, "Recent Work on the Principles of Mathematics," *International Monthly*, Vol. 4, 1901, pp. 83-101. Reprinted in "The Collected Papers of Bertrand Russell Vol. 3*,*" G.H. Moore (ed.), Routledge, London, 1993, pp. 366-379, at p. 366. I owe this reference to G. Bornet's section on "Frege's psychologism criticism (of Boole)," in "George Boole: Selected Manuscripts on Logic and Its Philosophy," I. Grattan-Guinness and G. Bornet (ed.),Science Networks Historical Studies 20, Birkhäuser Verlag, Basel, Boston, and Berlin, 1997, pp. xlviii-l.

[17] G. Boole, "The Mathematical Analysis of Logic" [see note 6], p. 47.

[18] G. Boole, "An Investigation of the Laws of Thought" [see note 7], p. 1.

[19] For such an attempt, see G. Bornet, in "George Boole: Selected Manuscripts" [see note 13], lxiv, note 22.

[20] G. Boole, "An Investigation of the Laws of Thought" [see note 7 above], p. 11.

[21] G. Boole, "The Mathematical Analysis of Logic" [see note 6], p. 52.

[22] G. Boole, "An Investigation of the Laws of Thought" [see note 7 above], p. 11.